\documentclass[12pt]{article}
\usepackage{eurosym}
\usepackage[left=0.8in, right=0.8in, top=0.8in, bottom=0.8in]{geometry}
\usepackage{hyperref}
\usepackage{mathtools}
\usepackage[]{lipsum}
\usepackage{blindtext}
\usepackage{graphicx}
\usepackage{mathrsfs}
\usepackage{graphicx}
\usepackage{caption}

\usepackage{authblk}
\providecommand{\keywords}[1]{\textit{keywords: } #1}
\usepackage{authblk}
\title{Quantum memory assisted entropic uncertainty and entanglement dynamics: Two qubits coupled with local fields and Ornstein Uhlenbeck noise}
\author[1] {{Atta Ur Rahman}\footnote{Email: Attapk@outlook.com}}
\author[2] {Nour Zidan}
\author[3] {S. M. Zangi}
\author[4] {Hazrat Ali}
\affil[1]{Key Laboratory of Aerospace Information Security and Trusted Computing, School of Cyber Science and Engineering, Wuhan University, P.O. Box 430072, Wuhan, China}
\affil[2]{Mathematics Department, College of Science, Jouf University, P.O. Box 2014 Sakaka, Saudi Arabia}
\affil[3]{School of Physics and Astronomy, Yunnan University, P.O. Box 650500 Kunming, China}
\affil[4]{Abbottabad University of Science and technology, P.O. Box 22500 Havellian KP, Pakistan}
\date{}                     
\begin{document}

\maketitle

\begin{abstract}
Entropic uncertainty and entanglement are two distinct aspects of quantum mechanical procedures. To estimate entropic uncertainty relations, entropies are used: the greater the entropy bound, the less effective the quantum operations and entanglement are. In this regard, we analyze the entropic uncertainty, entropic uncertainty lower bound, and concurrence dynamics in two non-interacting qubits. The exposure of two qubits is studied in two different qubit-noise configurations, namely, common qubit-noise and independent qubit-noise interactions. To include the noisy effects of the local external fields, a Gaussian Ornstein Uhlenbeck process is considered. We show that the rise in entropic uncertainty gives rise to the disentanglement in the two-qubit Werner type state and both are directly proportional. Depending on the parameters adjustment and the number of environments coupled, different classical environments have varying capacities to induce entropic uncertainty and disentanglement in quantum systems. The entanglement is shown to be vulnerable to current external fields; however, by employing the ideal parameter ranges we provided, prolonged entanglement retention while preventing entropic uncertainty growth can be achieved. Besides, we have also analyzed the intrinsic behavior of the classical fields towards two-qubit entanglement without any imperfection with respect to different parameters.
\end{abstract}
\maketitle
\section{Introduction}
In quantum physics, Heisenberg's uncertainty principle is a fundamental concept. Uncertainty relations in terms of entropies were constructed to solve conceptual inadequacies in the uncertainty principle's original formulation, and they now play a crucial role in quantum foundations \cite{1}. In the security analysis of some quantum systems, entropic uncertainty relations have lately emerged as a significant component \cite{2}. The uncertainty principle in quantum mechanics is both a fundamental feature and a significant departure from classical physics. Any pair of incompatible observables obeys a specific type of uncertainty relationship, imposing final constraints on measurement accuracy while also laying the theoretical groundwork for future technologies, such as quantum encryption and quantum information \cite{3,4,5,6}. The newly empirically validated entropic uncertainty principle has piqued interest in its potential applications from various perspectives. A new sort of Heisenberg relation known as the quantum memory assisted entropic uncertainty relation has just been constructed, according to Renes and Boileau's concept \cite{7,8}. The entropic uncertainty relation is used in cryptographic security \cite{9}, quantum randomness \cite{10}, quantum key distribution \cite{11}, probing quantum correlations \cite{12}, entanglement witnessing \cite{13}, and quantum metrology \cite{14}.

Quantum entanglement is a quantum mechanical phenomenon in which the quantum states of two or more objects, notwithstanding their spatial separation, must be explained in terms of one another \cite{15}. As a result, there are correlations between the systems' observable physical characteristics. Even though quantum physics makes it difficult to forecast which set of measurements will be observed, it is feasible to combine two particles into a single quantum state so that when one is detected as spin-up, the other is always detected as spin-down, and vice versa. Quantum entanglement has been utilized in experiments to establish quantum teleportation \cite{16}, and it has potential applications in quantum computing \cite{17}, quantum cryptography \cite{18}, communications \cite{19}, quantum radar \cite{20} and entanglement swapping \cite{21}.

The preservation of entanglement and the level of uncertainty in open quantum systems are inextricably linked, and a hotly debated topic. The entropic uncertainty may give rise to different phenomena such as rise in entropy, mixedness and dephasing of the systems. These effects can further cause the state to be disentangled. We are motivated by these possible reasons to link entropic uncertainty and entanglement in two entangled qubits. In this case, we are interested in analyzing the dynamical map of a Werner type mixed entangled state and the relationship between entanglement and entropic uncertainty and the related optimal control. In addition, this is significant because the dynamics of open quantum systems are crucial for the development of quantum protocols and the inter-transmission of information between two locations \cite{21}. The principal source of uncertainty is that quantum systems cannot be completely isolated from their external mediums, which can accommodate a variety of disorders \cite{22, 23, 24, 25}. These disorders generate a variety of noises, which, when superimposed on the phase factors of the systems, reduce the efficiency of quantum processes and phenomena \cite{26,27,28}. As a result, research into such topics can help to reduce the actual causes of quantum mechanical application failure while also improving relative precision and measurement accuracy by reducing and optimally controlling the entropic uncertainty and hence entanglement in quantum systems.

In the present work, we discuss the dynamics of entropic uncertainty, entropic uncertainty bound, their relative difference and entanglement dynamics in a two-qubit mixed entangled state under the influence of classical fields. To limit our problem, we consider Ornstein Uhlenbeck (OU) noise generation in the classical fields and the main reason for the disentanglement of the two qubits and the relative degree of entropic uncertainty. In microscopic view, the OU noise is caused by the Brownian motion of the particles, which can be found nearly in every quantum mechanical process \cite{28}. This makes our noise model more significant because of its widespread presence and noisy actions in such operations. We prefer the classical context of environments rather than the non-classical ones because the local former provides more degrees of freedom to examine the dynamical maps of quantum systems \cite{22, 23, 24}. Two types of two-qubit spin squeezing models were used to investigate thermal quantum correlations and the entropic uncertainty relation in the presence of quantum memory \cite{29}. For two atoms and the relative dynamics of the entanglement and uncertainty was found to be greatly dependent upon the temperature parameters \cite{30}. For different three levels systems, the dynamics of entropic uncertainty reveal that the corresponding entanglement losses and rise in entropy and uncertainty is regulated by the coupling strengths of the random telegraphs noise \cite{31}. The demonstration of the evolution of entropic uncertainty in the multi-measurement process has shown that the Markovianity and non-Markovianity of the fields can be traced back to the degree of uncertainty and noise \cite{32}. A new type of long-range reaction was used to achieve long-distance entanglement in the spin system \cite{33, 34}. In a similar case, the authors in \cite{36} investigated the dynamics of entropic uncertainty in three qubits and they found that the classical depolarizing noise and environments enhance entropy. Besides, they found that designing the system-environment coupling between three qubits and environments can also lead to enhanced entanglement preservation and lower entropic uncertainty. Thus, the above literature concludes that entropic uncertainty and entanglement are controlled by the different variables and fields, which must be thoroughly investigated for practical implementation of the quantum protocols.

We assume two kinds of system-environment coupling: common qubit-noise (CQN) and independent qubit-noise (IQN) configurations. Both qubits will be coupled to a common environment characterized by an OU noise source in the CQN environment. The two qubits are coupled with two independent local environments in the IQN configuration case. This will help to conclude the entropy and disentanglement level for the increasing number of environments. Entanglement has already been shown to degrade differently in different types and number of environments \cite{22,23,24,25,26,27,28}.

This paper is organized as: In the Sec.\ref{Model}, we give the details of the physical model, estimators of entropy and entanglement and OU noise application. The explicit results and discussion are written in Sec.\ref{Results}. In Sec.\ref{Conclusion}, we summarize our results in few remarks.  

\section{Suggested model and dynamics}\label{Model}
Our model comprises two non-interacting qubits initially prepared in a mixed entangled state coupled with a classical environment. OU noise is considered as being the primary cause of dephasing and entropic increase in classical environments. We examine CQN and IQN configurations, which are two different designs of system-environment coupling approaches. In CQN configuration, the dynamical map of the two qubits is studied under the influence of a single OU noise source. In IQN configuration, the system is considered evolving under the influence of two independent OU noise sources. The Hamiltonian, which characterizes the current model, is written as;

\begin{figure}[!h]
	\begin{center}
		\includegraphics[width=0.49\textwidth, height=160px]{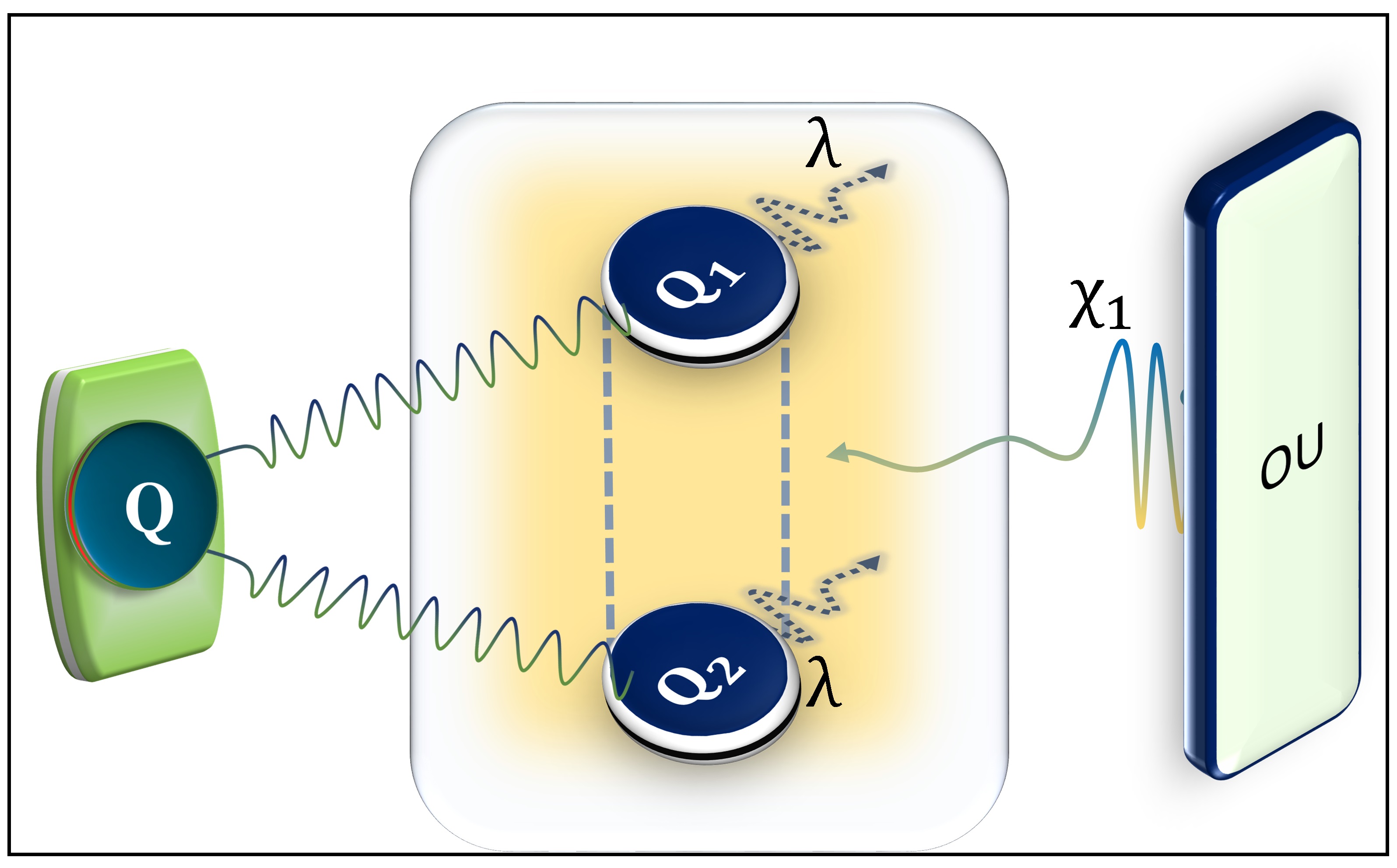}
		\put(-200,165){($ a $)} \ 
		\includegraphics[width=0.47\textwidth, height=160px]{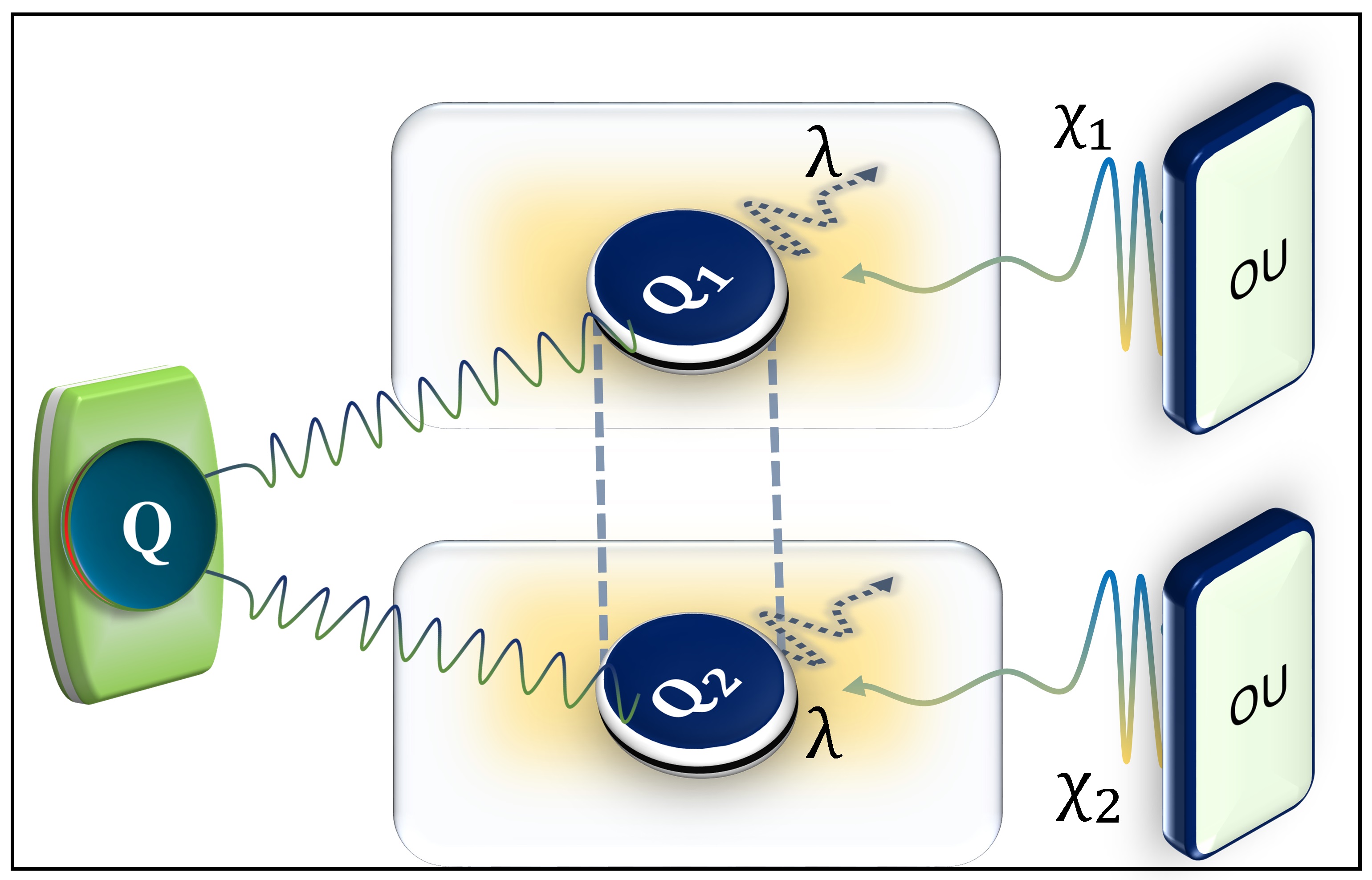}
		\put(-200,165){($ b $)}
		\end{center}

	\caption{Shows the schematic diagram of two qubits $Q_1$ and $Q_2$ connected with square-like boxes means classical environments of two types: common qubit-noise configuration (a), and independent qubit-noise configuration (right) with a linked quantum memory $Q$ for utilizing the concept quantum memory assisted entropic relations. OU denotes the Ornstein Uhlenbeck noisy sources and the yellowish-blue lines represent the connection between the classical channels and noisy sources characterized by the stochastic parameter $\chi_n(t)$. The glow around the qubits shows the entropic action of the environments while, the wavy lines above the qubits represent the dynamics and the relative coupling strengths $\lambda$ and its diminishing size shows the resultant dephasing effects.}
	\end{figure}	

\begin{align}
H(t)=H_1(t) \otimes I_2+I_1\otimes H_2(t),
\end{align}
where $H_n(t)=\kappa I+\lambda \chi_n(t) \sigma_z$ with $n \{1,2\}$. $\kappa$ represents the energy of the relative qubit, $I$ and $\sigma_z$ are the identity and Pauli matrices of dephasing classical channels while $\lambda$ is the coupling constant, regulating the strength of linking between the qubits and classical environments. The terms $\chi_n(t)$ represent the stochastic parameter of the fields and control the flipping of the qubits between $\pm1$. For the time-evolution of the two qubits in classical fields, we employ the time unitary operation by:
\begin{align}
U(t)=\exp[-i \int^t_{0} H(f)df].\label{time unitory matrix}
\end{align}
Time evolved state of the two qubits, when prepared in the initial state $
\rho_0$ is obtained using:
\begin{align}
\rho(t)=U(t)\rho_0 U^{\dagger}(t).\label{time evolved density matrix}
\end{align}
We assume the OU process, a stochastic mathematical process with applications in both physical sciences and finance, impacts the classical fields to account for noise. This term in physics describes the velocity of a massive Brownian particle under the influence of friction. In many quantum mechanical protocols, the OU method is a static Gaussian–Markov operation with OU noise, and it has been identified as one of the numerous and primary causes of information, coherence, and quantum correlation losses \cite{28}. OU noise has been extensively studied in the case of single qutrit, two-qubit, three qubits, and hybrid qubit-qutrit states and we find that in each case, the degree and behaviour of losses are different \cite{26, 37, 38}. Currently, the OU noise is applied to the dynamical map of the two-qubit mixed Werner state, which is created in the state $\rho_0$. To determine the negative consequences of OU noise, we use a zero-mean Gaussian process ($\langle \mathcal{G}(t) \rangle=0$) to describe the classical field $\mathcal{L}(t)$ affecting the system. This is further defined by the auto-correlation function, and has the form:
\begin{align}
A(g, t-t^{\prime})=\dfrac{\exp[-g|t-t^{\prime}|]g }{2}.\label{KROU}
\end{align}
We connect the classical noise with the environments in the dynamical map of the two-qubit state using the $\beta$-function, which is written as \cite{28}:
\begin{equation}
\beta_{OU}(t)=\int_0^t \int_0^t A(s-s)^{\prime}ds ds^\prime. \label{Beta function}
\end{equation}
We can extract the final $\beta$-function for the OU noise by plugging the auto-correlation function from Eq.\eqref{KROU} into Eq.\eqref{Beta function} as:
\begin{equation}
\beta_{{OU}}(t)=\frac{1}{g}[g \tau+\exp[-g \tau]-1],\label{Beta function of OU}
\end{equation}
The memory characteristic of the classical environment is controlled by $g$ and for the $OU$ noise case, we consider $t=\tau$.
\subsection{The entropic uncertainty measure}
Consider the following network with two users, Bob and Alice: Bob generates a qubit in the quantum state of his choice and delivers it to Alice, who must choose between the two measurements and broadcast her decision to Bob. Now, we can reduce the uncertainty in the outcome using the measurement results obtained by Bob. The standard deviation uncertainty relation for two observables $A$ and $B$ can be written as \cite{39}:
\begin{align*}
\Delta A \Delta B \geq \frac{1}{2} |\langle[A , B] \rangle|.
\end{align*}
Deutsch proposed the entropic uncertainty relation for any pair of observables by describing uncertainty in terms of Shannon entropy rather than standard deviation \cite{40}. Maassen and Uffink devised a tighter entropic uncertainty expression based on Deutsch's approach \cite{41}:
\begin{align}
S(A)+S(B)\geq \log_2(\frac{1}{c}),
\end{align}
where $S(A)$ is the Shannon entropy, which represents the probability distribution when A is measured, and $S(B)$ is the Shannon entropy when B is measured. $c$ denotes the complementary of $A$ and $B$, and $c=\max_{a,b}\vert \langle \psi \vert \phi \rangle \vert^2$ for non-degenerate observables, where $\vert \psi\rangle$ and $|\phi\rangle$ are the eigenvectors of $A$ and $B$, respectively. The current definition has been updated into a new form known as a quantum memory assisted entropic uncertainty relation \cite{42}, which has also been experimentally tested \cite{43} and can be written as:
\begin{align}
S(A\vert 2) + S(B\vert 2) \geq S(1\vert 2)+\log_2(\frac{1}{c}),\label{ULR}
\end{align}
where $S(1\vert 2)=S(\rho_{12})-S(\rho_2)$ is the conditional von-Neumann entropy. In Eq.\eqref{ULR}, $R(\tau)$ and $L(\tau)$ represents the left and right-hand sides. To find the difference between the two sides, we use the equation:
\begin{equation}
U(\tau)= L(\tau)-R(\tau)\label{tightness}
\end{equation}
as the tightness of the uncertainty relation. Once the first qubit is measured by $A$, the system's post-measurement state can be stated as \cite{44}:
\begin{align}
\rho_{A2}=\sum_n (\vert \psi_n \rangle_1 \langle \psi_n \vert \otimes I_2) \rho_{12}(\vert \psi_n\rangle_1 \langle \psi_n \vert \otimes I_2),
\end{align}
Note that we utilize the Werner state form of the two non-interacting entangled qubits:
\begin{align}
\rho_0=\frac{1-p}{4} (I_4)+ p\vert \psi \rangle \langle \psi \vert, \label{Initial denisty matrix}
\end{align}
where $\vert \psi \rangle=\frac{1}{\sqrt{2}}(\vert 00 \rangle+\vert 11 \rangle) $ is the two qubit maximally entangled Bell's state, $p$ denotes the purity factor, controlling the initial purity in the system and ranges between $0 \leq p \leq 1$.
\subsection{Concurrence}
To assess entanglement, we use concurrence for the bipartite state, which ranges from $1 \geq C(t) \geq 0 $. The state is entangled at $C(t)=1$, but at the lowest bound, the state becomes completely separable. For the two-qubit state, the concurrence measurement can be carried out using the following expression \cite{15, 25}:
\begin{equation}
C= \max \{0,\sqrt{\nu_1}-\sqrt{\nu_2}-\sqrt{\nu_3}-\sqrt{\nu_4} \}, \label{concurrence}
\end{equation}
where $\nu_i$ are the eigenvalues of the time evolved density matrix $\rho(t)$ in decreasing order.
\section{Analytical results}\label{Results}
In this section, we present the entanglement and entropic uncertainty dynamics results obtained using Eq.\eqref{tightness} and \eqref{concurrence}. The system's time unitary matrix, obtained using Eq.\eqref{time unitory matrix}, has the following form:
\begin{equation}
U(t)=\left(
\begin{array}{cccc}
 e^{i t (-2 \kappa +(  \chi_a(t)+  \chi_b(t)) \lambda )} & 0 & 0 & 0 \\
 0 & e^{-i t (2 \kappa +(-  \chi_a(t)+  \chi_b(t)) \lambda )} & 0 & 0 \\
 0 & 0 & e^{-i t (2 \kappa +(  \chi_a(t)-  \chi_b(t)) \lambda )} & 0 \\
 0 & 0 & 0 & e^{-i t (2 \kappa +(  \chi_a(t)+  \chi_b(t)) \lambda )}
\end{array}
\right)
\end{equation}
\begin{figure}[!h]
	\begin{center}
		\includegraphics[width=0.47\textwidth, height=160px]{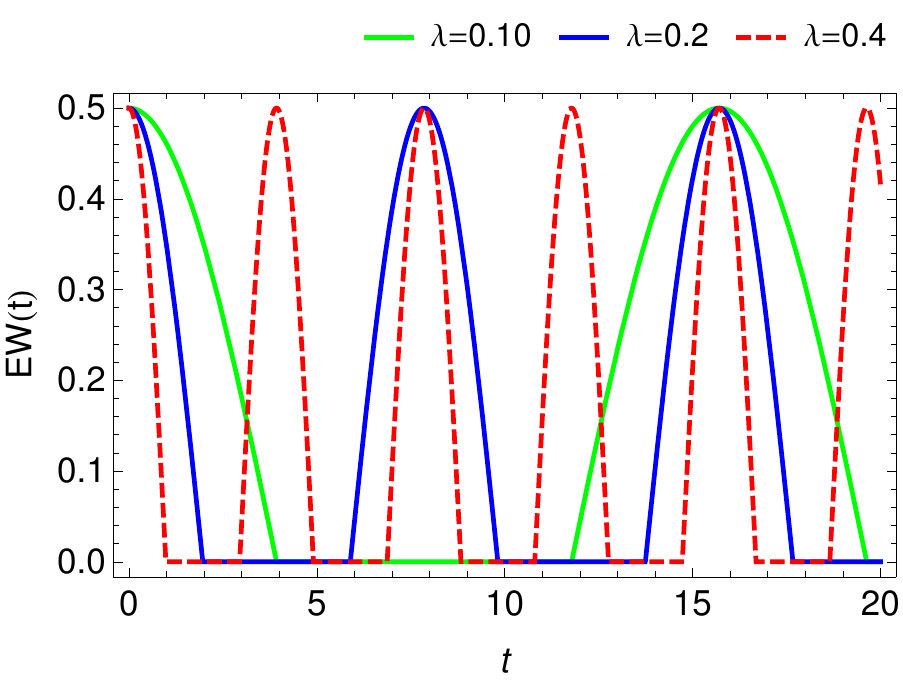}
		\put(-200,165){($ a $)} \ 
		\includegraphics[width=0.47\textwidth, height=160px]{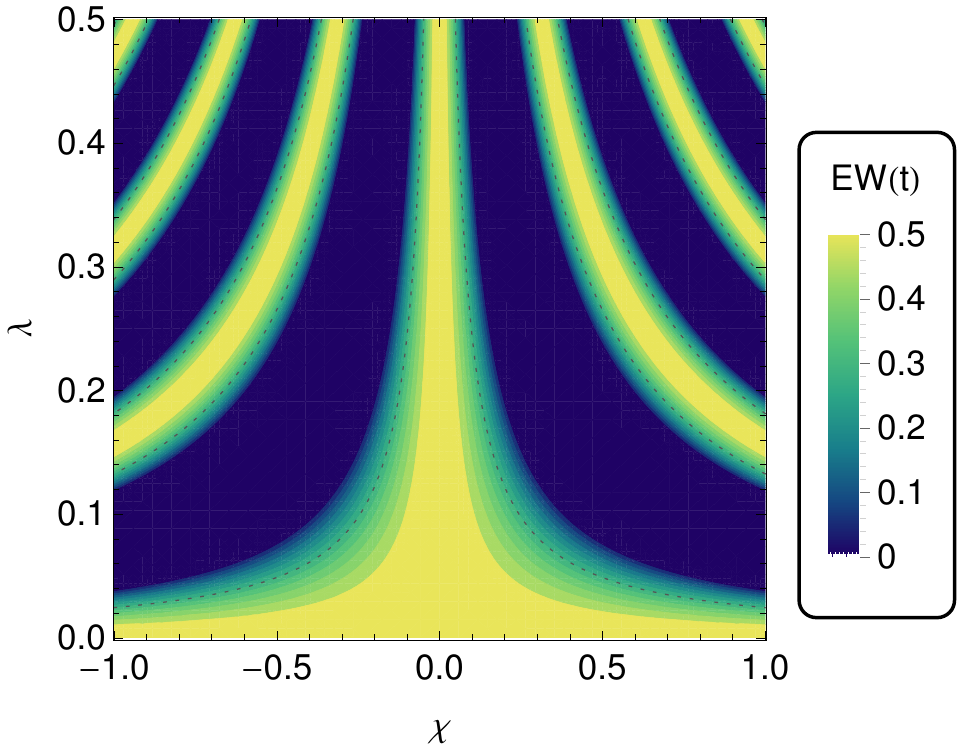}
		\put(-200,165){($ b $)} \\ 
		\includegraphics[width=0.47\textwidth, height=160px]{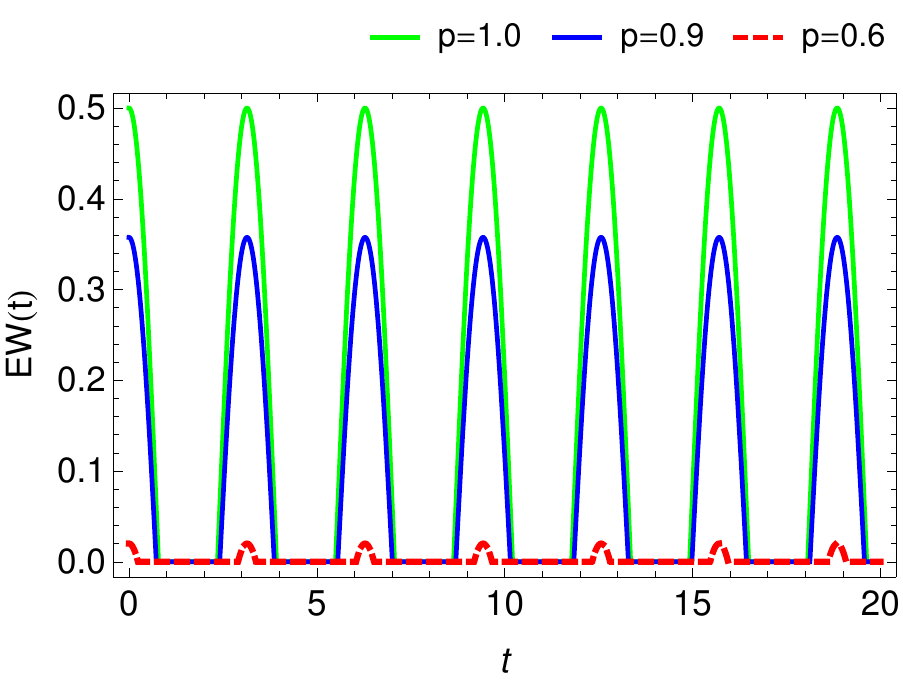}
		\put(-200,165){($ c $)}\
		\includegraphics[width=0.47\textwidth, height=160px]{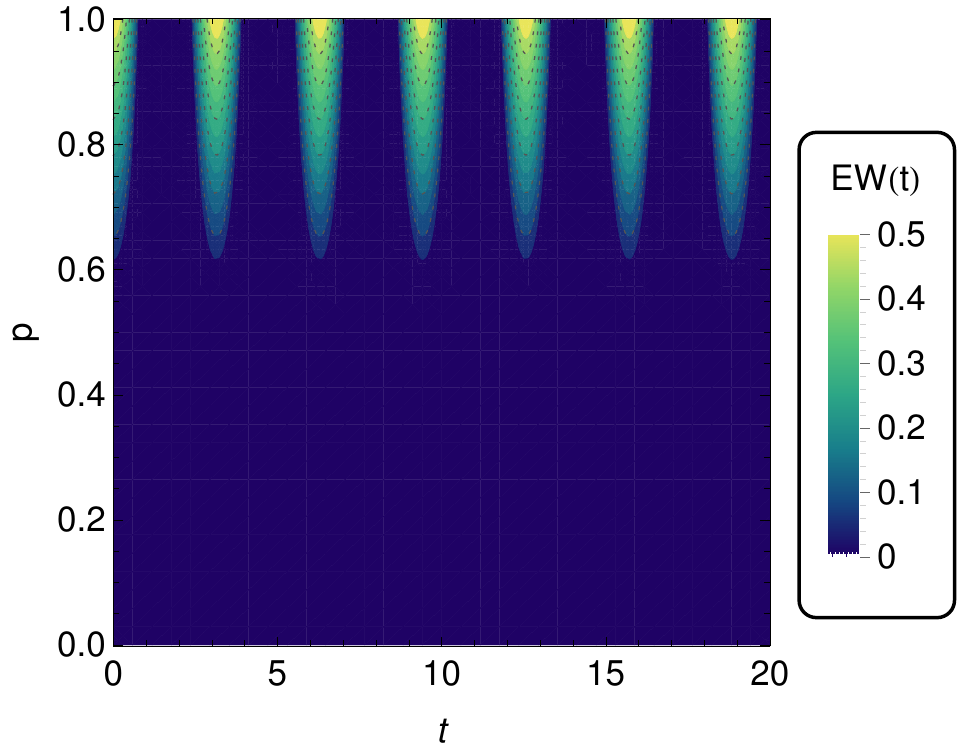}
		\put(-200,165){($ d $)} 
		\end{center}
\caption{Dynamics of $EW$ for varying values of $\lambda$ (a), $\lambda~\text{vs}~\chi$ (b), for varying values of $p$ (c) and $p~\text{vs}~t$ (d) in two-qubit state, $\vert \psi \rangle=\frac{1}{\sqrt{2}}(\vert 00 \rangle+\vert 11 \rangle)$ initially prepared in the state $\rho_0$ given in Eq.\eqref{Initial denisty matrix} in common qubit-noise configuration.}\label{EWO}
\end{figure}

Note that all operations are carried out using the time evolved density matrix given in Eq.\eqref{time evolved density matrix} which has the following form:
\begin{align}
\rho_{CQN}(t)=\left(
\begin{array}{cccc}
 \frac{1+p}{4} & 0 & 0 & \frac{1}{2} e^{4 i t   \chi_a(t) \lambda } p \\
 0 & \frac{1-p}{4} & 0 & 0 \\
 0 & 0 & \frac{1-p}{4} & 0 \\
 \frac{1}{2} e^{-4 i t   \chi_a(t) \lambda } p & 0 & 0 & \frac{1+p}{4}
\end{array}
\right)\label{time evolved density ensemble}
\end{align}
Fig.\ref{EWO} shows the time evolution of entanglement detection in two qubits prepared in the state $\rho_0$. Because the values of these parameters are equal i.e., $\pm1$, we set $\chi_a=\chi_b=\chi$ in this section. Compared to those with defects, understanding the intrinsic behaviour of classical environments without defects is critical for promoting and preserving non-local correlation. As a result, we employ entanglement witness ($EW$), a tool for detecting entanglement. Mathematically, $EW=-Tr[\rho(t).E_x]$ is a simple estimable measure. Where, $\rho(t)$ is the time evolved density ensemble given in Eq.\eqref{time evolved density ensemble}, and $E_x=\frac{1}{2}I-\rho_0$ is the $EW$ operator. As shown in Fig.\ref{EWO}, the two qubits maintain their entanglement while experiencing revivals that show non-Markovian behaviour. Using Fig.\ref{EWO}(a), we investigated the effect of increasing the intensity of $\lambda$ on entanglement revivals. When $\lambda$ was increased, the robustness of the entanglement revivals improved. The results in Fig.\ref{EWO}(b) are consistent with those in Fig.\ref{EWO}(a), implying that as $\lambda$ increases, the revival speed increases. Meanwhile, the parameter $\chi$, which can be set between $\pm1$ and $\lambda$, simply toggles the system between relative maximum and minimum. We discovered that the two-qubit state entanglement is only effective in a narrow range, such as $0.6 < p \leq 1$, by plotting several values of the purity parameter $p$ against $t$ in Fig.\ref{EWO}(c). The results of Fig.\ref{EWO}(d) match those of Fig.\ref{EWO}(c), revealing the same $p$ region where the system remains entangled. It's worth noting that the two qubits and their surrounding classical environment effectively exchange information regularly, demonstrating that classical channels are vital resources for quantum information science applications. The behaviour of time evolved density ensemble in common and independent coupling cases will remain the same, as when no noise is involved, we get $\chi_a=\chi_b=\chi$.
\subsection{Entropic uncertainty relations and entanglement dynamics in common qubit-noise configuration}
We discuss the dynamics of entropic uncertainty, tightness and entanglement when two non-interacting qubits are both coupled with a common OU noise source in a single local random field. To include the OU noisy effects in the matrix given by Eq.\eqref{time evolved density matrix}, we take average of the time evolved density matrix of the system for the CQN configuration as follows:
\begin{align}
\rho_{CQN}(\tau)=\langle \rho(\phi_1, t)\rangle_{\theta_1},\label{CQN}
\end{align}
where $\phi_1$ is the combined factor of system and environments while $\theta_1$ is the superimposed noise phase over the system. In Eq.\eqref{CQN}, $\phi_1=\iota n \chi_1(t)$ and we set $\chi_1=\chi_2$ where $\theta=-\frac{1}{2} n^2 \beta(\tau)$. The explicit form of the Eq\eqref{CQN} obtained can be put into the following form:
\begin{equation}
\rho_{CQN}(\tau)=\left(
\begin{array}{cccc}
 \frac{1+p}{4} & 0 & 0 & \frac{e^{-2 \beta}}{2} \\
 0 & \frac{1-p}{4} & 0 & 0 \\
 0 & 0 & \frac{1-p}{4} & 0 \\
 \frac{e^{-2 \beta}}{2} & 0 & 0 & \frac{1+p}{4}
\end{array}
\right)
\end{equation}
where $\beta$-function is given in Eq.\eqref{Beta function}. The analtyical results of the Eq.\eqref{tightness} and Eq.\eqref{concurrence} takes the form as:
\begin{align}
U(\tau)=&\frac{1}{\text{Log}[16]}e^{-\frac{1}{2} n^2 \beta} \left(\text{T1}+e^{\frac{1}{2} n^2 \beta} (\text{T2}+\text{T3})\right),\\
C(\tau)=&-\sqrt{1-p}-\frac{1}{2} \text{T4}+\frac{1}{2} \text{T5},
\end{align}
where,
\begin{align*}
T1=&-4 \text{ArcTanh}\left[e^{-\frac{1}{2} n^2 \beta}\right]-2 \text{Log}\left[1-2 e^{-\frac{1}{2} n^2 \beta}+p\right]+2 \text{Log}\left[1+2 e^{-\frac{1}{2} n^2 \beta}+p\right],\\
T2=&-\text{Log}[16]-2 \text{Log}\left[\frac{1}{4}-\frac{1}{4} e^{-\frac{1}{2} n^2 \beta}\right]-2 \text{Log}\left[1+e^{-\frac{1}{2} n^2 \beta}\right]-2 (1+p) \text{Log}[1+p],\\
T3=&(1+p) \text{Log}\left[1-2 e^{-\frac{1}{2} n^2 \beta}+p\right]+(1+p) \text{Log}\left[1+2 e^{-\frac{1}{2} n^2 \beta}+p\right],\\
T4=&\sqrt{e^{-\frac{1}{2} n^2 \beta} \left(-2+e^{\frac{1}{2} n^2 \beta}+e^{\frac{1}{2} n^2 \beta} p\right)},\\
T5=&\sqrt{e^{-\frac{1}{2} n^2 \beta} \left(2+e^{\frac{1}{2} n^2 \beta}+e^{\frac{1}{2} n^2 \beta} p\right)}.
\end{align*}

\begin{figure}[!h]
	\begin{center}
		\includegraphics[width=0.47\textwidth, height=160px]{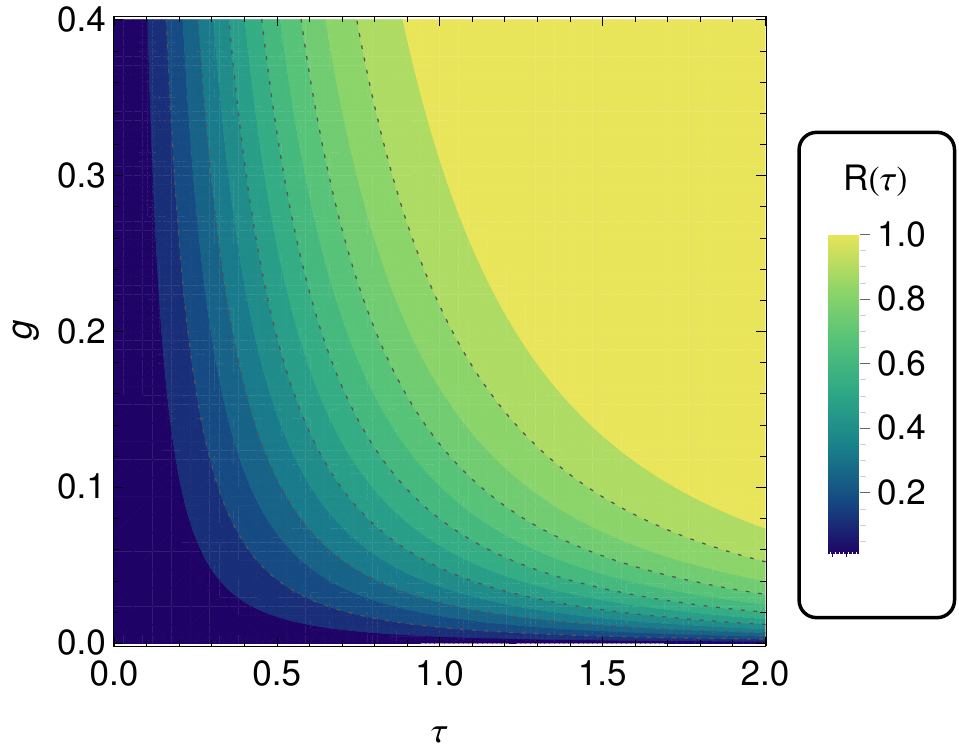}
		\put(-200,165){($ a $)} \ 
		\includegraphics[width=0.47\textwidth, height=160px]{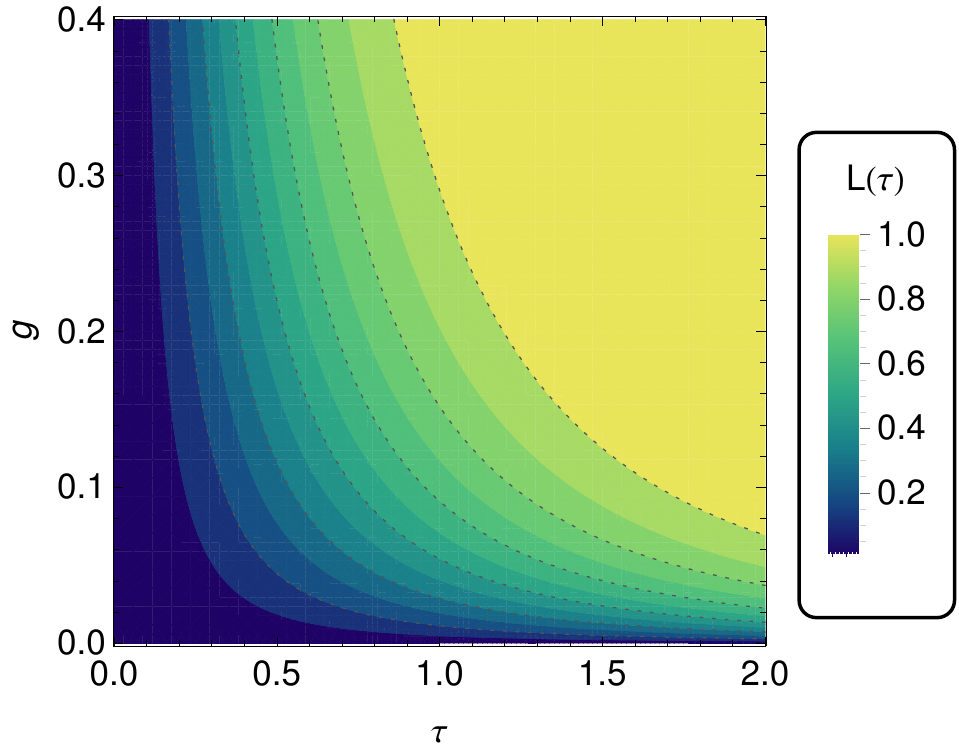}
		\put(-200,165){($ b $)} \\ 
		\includegraphics[width=0.47\textwidth, height=160px]{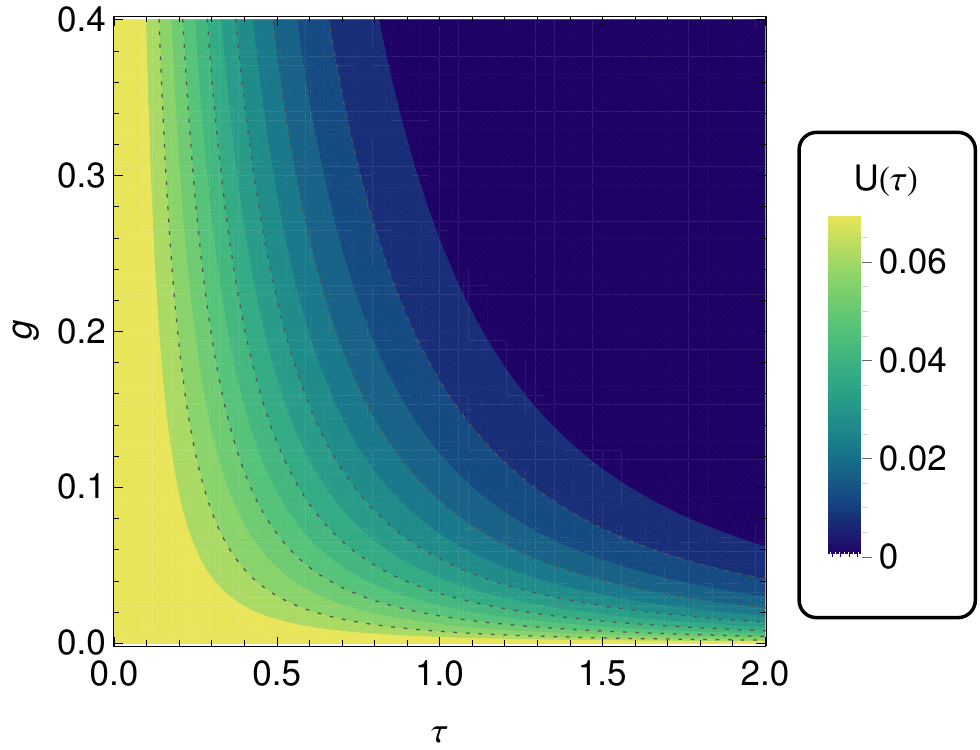}
		\put(-200,165){($ c $)}\
		\includegraphics[width=0.47\textwidth, height=160px]{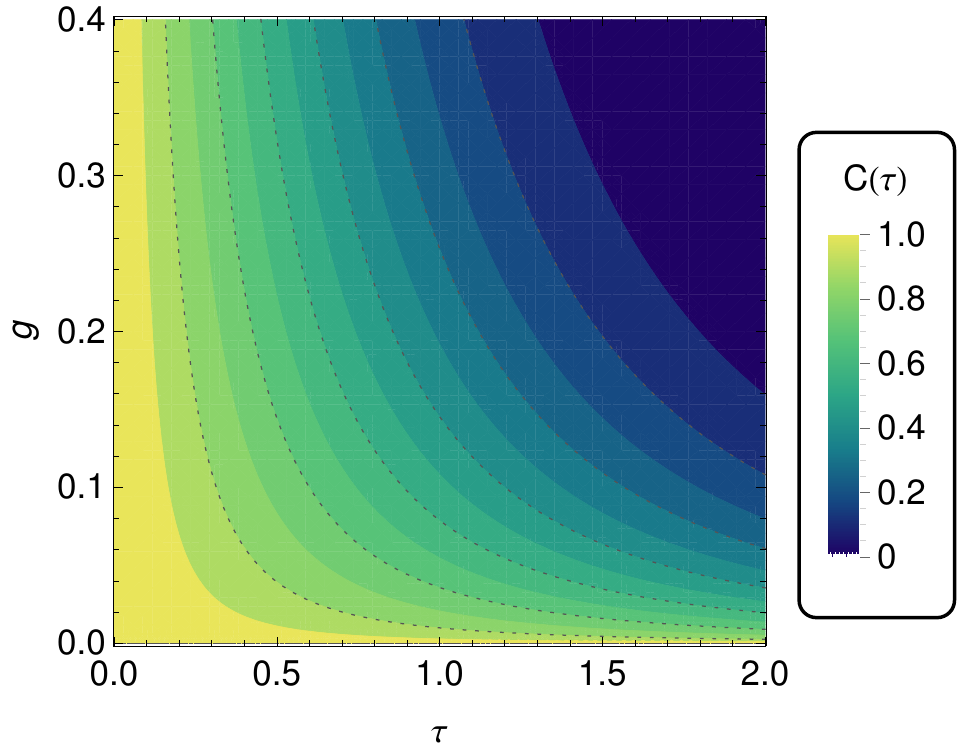}
		\put(-200,165){($ d $)} 
		\end{center}
\caption{Dynamics of $R(\tau)$ (a), $L(\tau)$ (b), $U(\tau)$ (c) and $C(\tau)$ (d) in two-qubit state, $\vert \psi \rangle=\frac{1}{\sqrt{2}}(\vert 00 \rangle+\vert 11 \rangle)$ initially prepared in the state $\rho_0$ given in Eq.\eqref{Initial denisty matrix} common qubit-noise configuration with parameter settings: $g=0.4$, $p=1$ against time parameter $\tau$.}\label{CQN-3D}
\end{figure}

When the CQN configuration is considered, Fig.\ref{CQN-3D} shows the entropic uncertainty, entropic uncertainty bound, tightness and concurrence dynamics utilizing Eqs.\eqref{ULR} and \eqref{concurrence} in the dynamical map of two qubits. When OU noise dephasing effects are present, entanglement decreases and entropic uncertainty relations increases in classical environments. In two qubits, the entropic uncertainty functions $L(\tau)$ and $R(\tau)$ are increasing, while the tightness and entanglement functions, $U(\tau)$ and $C(\tau)$ are found decreasing. Although, the difference between $L(\tau)$ and $R(\tau)$ is insignificant, however, the high rate of entropic uncertainty increase cannot be omitted. The results of $U(\tau)$, which show dynamics in a small restricted elevation, confirm the minimal difference between the $L(\tau)$ and $R(\tau)$. As a result, the $L(\tau)$ and $R(\tau)$ results are in good agreement with $U(\tau)$. From the $C(\tau)$ results, we can see that classical fields with Brownian motion disorders cause entanglement to degrade and entropic uncertainty to increase. It is simple to deduce that the rate of disentanglement lags the entropic uncertainty growth by comparing the entropic uncertainty and concurrence dynamics. This means that an increase in entropic uncertainty causes the disentanglement of the two qubits. Despite this, the noise parameter $g$ regulates entropic uncertainty and entanglement loss, and as $g$ rises, entropic uncertainty rises and entanglement falls. Under the current noise and parameter settings, the system becomes completely separable because of high entropic uncertainty. We find that the current results differ completely from those described in \cite{45}, where previous results showed revivals in $U(\tau)$, $R(\tau)$, and $L(\tau)$. The maximum values of the two sides of uncertainty and tightness do not match, which is important. The qualitative monotone behaviour and dynamical map of the two non-equivalent sides, on the other hand, were similar and depict that the entanglement and information loss is irreversible in the current example.
\subsubsection{Explicit dynamics of the two qubits in common qubit-noise configuration}
\begin{figure}[!h]
	\begin{center}
		\includegraphics[width=0.47\textwidth, height=160px]{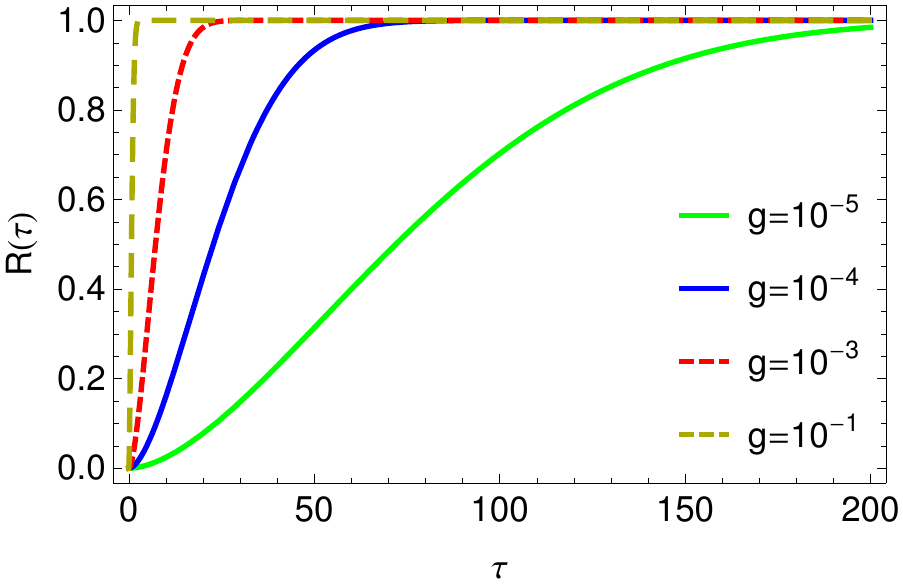}
		\put(-200,165){($ a $)} \ 
		\includegraphics[width=0.47\textwidth, height=160px]{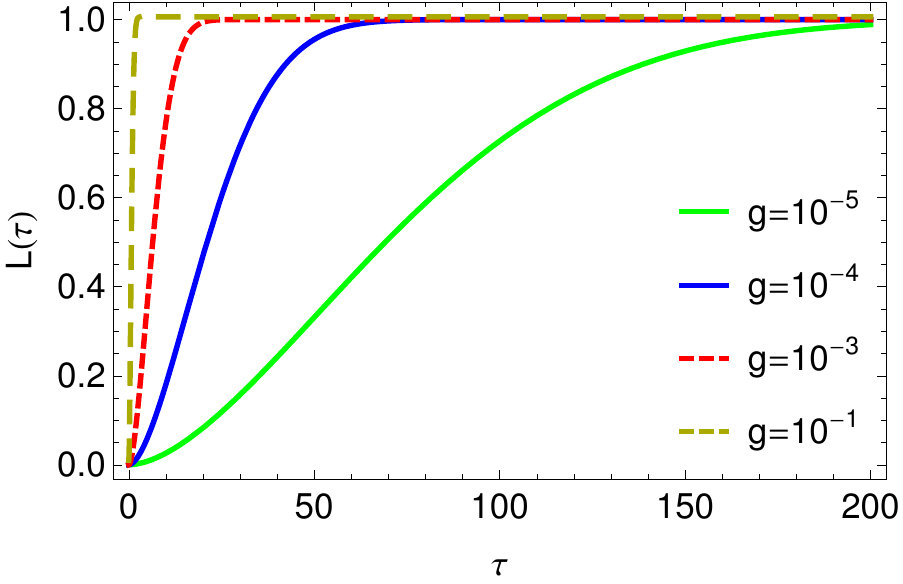}
		\put(-200,165){($ b $)} \\ 
		\includegraphics[width=0.47\textwidth, height=160px]{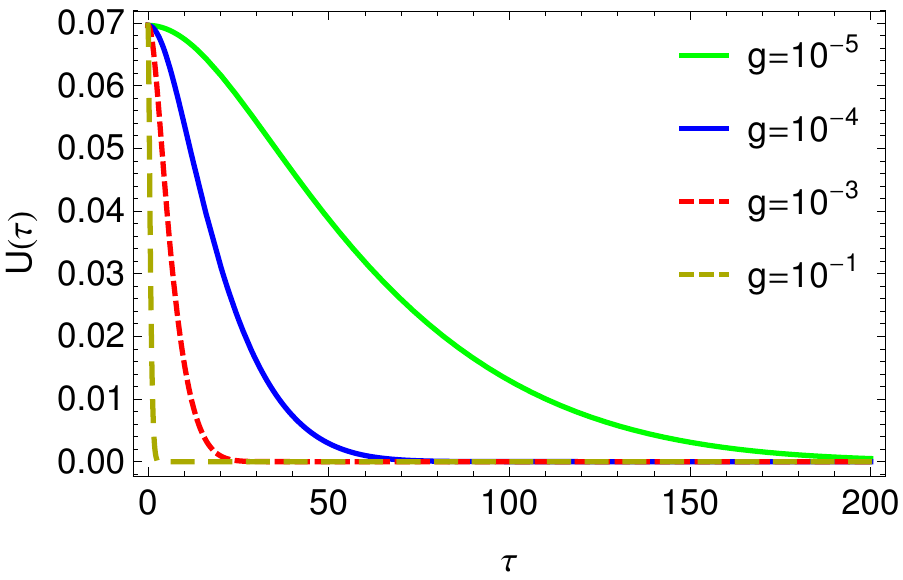}
		\put(-200,165){($ c $)}\
		\includegraphics[width=0.47\textwidth, height=160px]{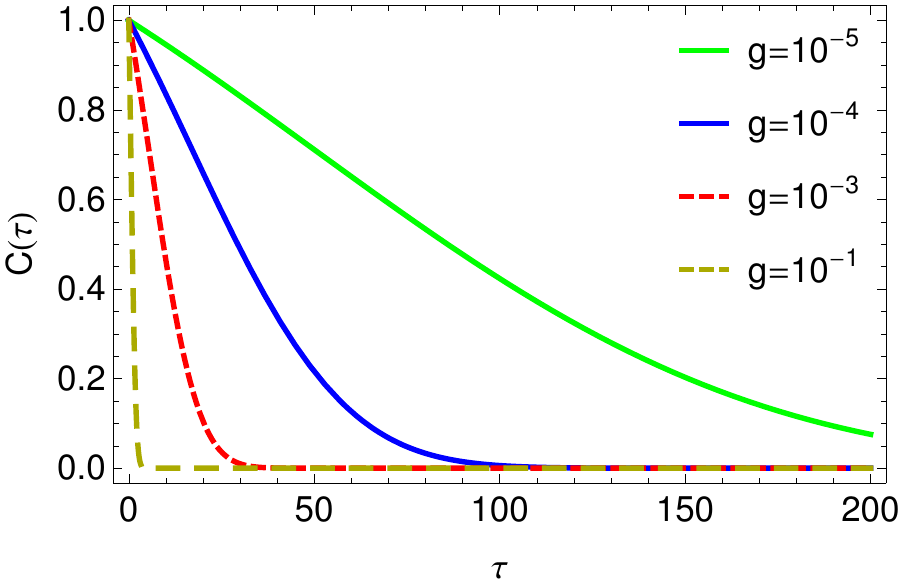}
		\put(-200,165){($ d $)} 
		\end{center}

\caption{Dynamics of $R(\tau)$ (a), $L(\tau)$ (b), $U(\tau)$ (c) and $C(\tau)$ (d) for different values of $g$ in two-qubit state, $\vert \psi \rangle=\frac{1}{\sqrt{2}}(\vert 00 \rangle+\vert 11 \rangle)$ initially prepared in the state $\rho_0$ given in Eq.\eqref{Initial denisty matrix} for common qubit-noise configuration when $p=1$ against time parameter $\tau$.}\label{different values of g}
\end{figure}

The entropic uncertainty, entropic uncertainty bound, tightness and concurrence dynamics are displayed in Fig.\ref{different values of g} when the system is coupled with a single classical environment. Initially, $R(\tau)=L(\tau)=U(\tau)=0$ and $C(\tau)=1$ suggesting the state has initially no uncertainty and is maximally entangled. When the interaction between the system and field is switched on, the entropic uncertainty rises and causes the entanglement to degrade. The current results are qualitatively similar to those in Fig.\ref{CQN-3D}, although they differ in quantitative terms. Lower $g$ values allowed the $L(\tau)$ and $R(\tau)$ to achieve ultimate saturation heights after the maximum entropic rise however taking a much longer time, which is the main reason for the disparity between the Figs.\ref{different values of g}, and \ref{CQN-3D}. The difference between the $L(\tau)$ and $R(\tau)$ is minor when compared to the results obtained for $g=0.4$ in Figs.\ref{CQN-3D} and \ref{different values of g}. This shows that entropic uncertainty is primarily caused and controlled by the noise parameter $g$ and that the two are inexorably related. The results of $U(\tau)$ show that $L(\tau) >R(\tau)$ and depict that the gap between the two sides of the uncertainty relation becomes narrow and finally vanishes with time. The entanglement decays monotonically under the influence of OU noise, as seen by the $C(\tau)$ measure. As the values of $g$ increase, the entanglement decreases. The findings of $C(\tau)$ suggest that rising entropic uncertainty directly affects the degree of entanglement between qubits and that entropic uncertainty increases faster than entanglement diminishes. This suggests that entropic uncertainty is a critical contributor in entangled quantum systems losing their entanglement. We discovered that for low $g$ values, we could maintain entanglement for a long time, even though the state eventually becomes separable. The current entropic uncertainty results contradict those reported in \cite{45, 46}, where the qualitative dynamics of entropic uncertainties, tightness, and entanglement dynamics are vastly different.

\begin{figure}[!h]
	\begin{center}
		\includegraphics[width=0.47\textwidth, height=160px]{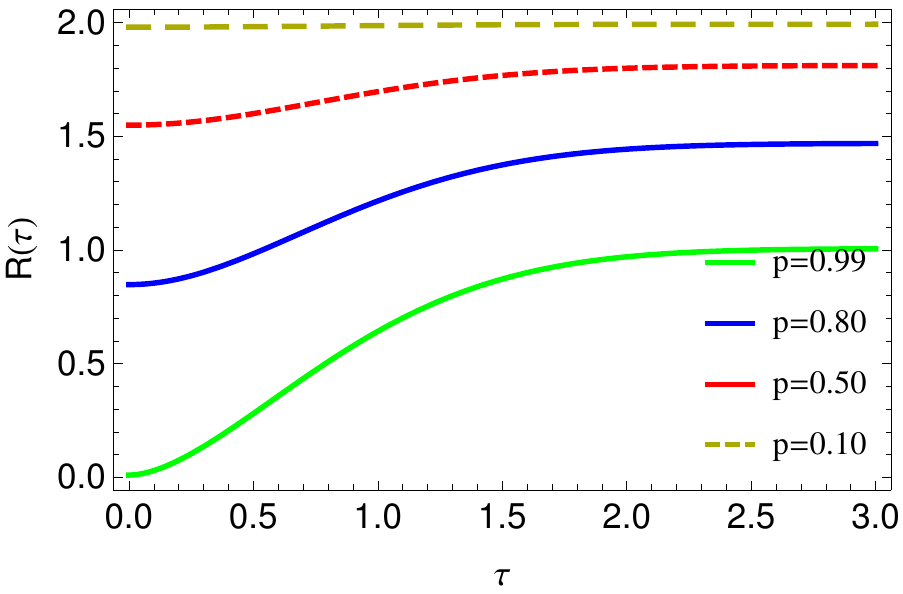}
		\put(-200,165){($ a $)} \ 
		\includegraphics[width=0.47\textwidth, height=160px]{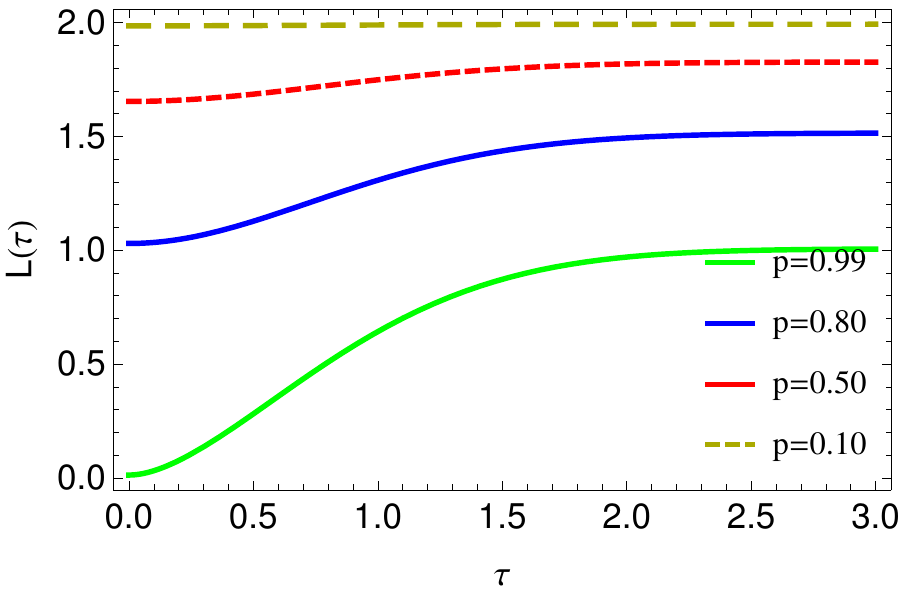}
		\put(-200,165){($ b $)} \\ 
		\includegraphics[width=0.47\textwidth, height=160px]{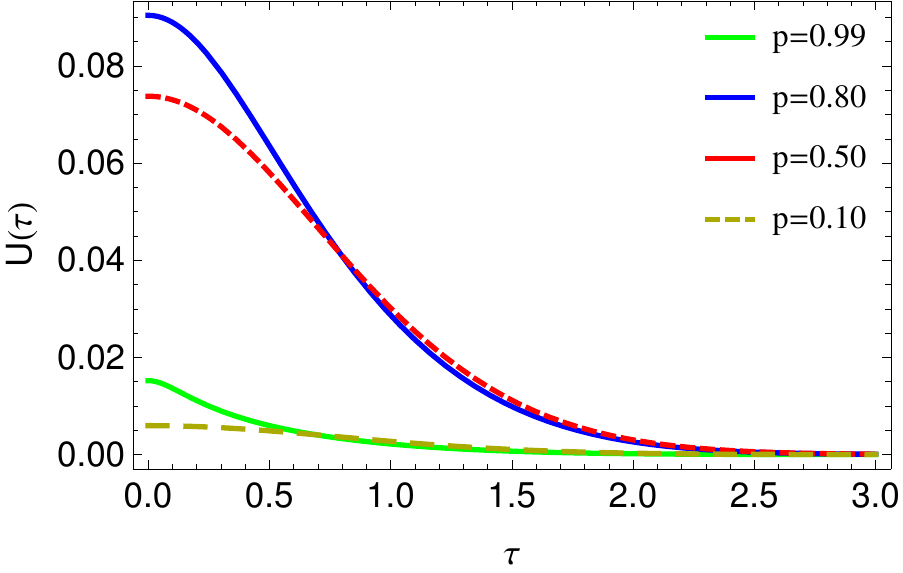}
		\put(-200,165){($ c $)}\
		\includegraphics[width=0.47\textwidth, height=160px]{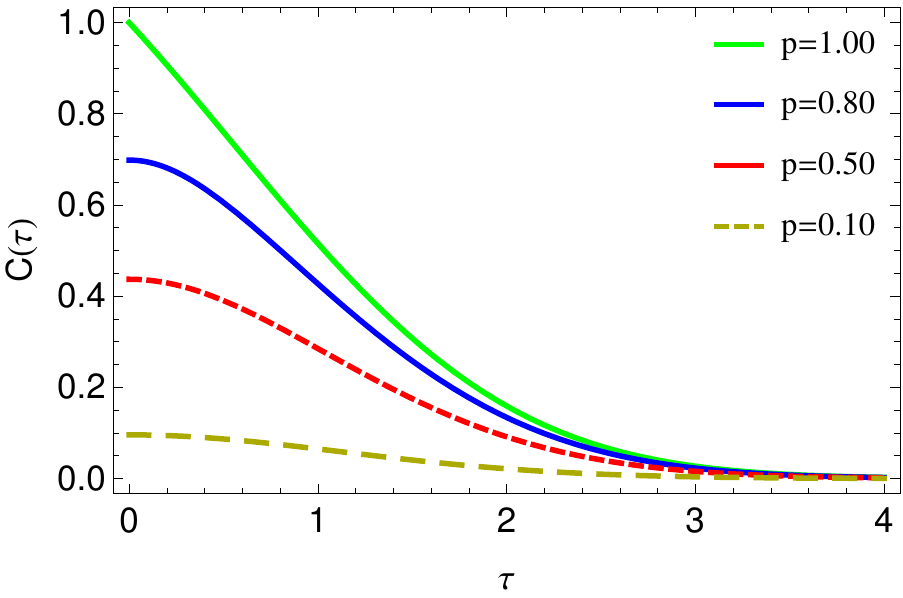}
		\put(-200,165){($ d $)} 
		\end{center}
\caption{Dynamics of $R(\tau)$ (a), $L(\tau)$ (b), $U(\tau)$ (c) and $C(\tau)$ (d) for different values of $g$ in two-qubit state, $\vert \psi \rangle=\frac{1}{\sqrt{2}}(\vert 00 \rangle+\vert 11 \rangle)$ initially prepared in the state $\rho_0$ given in Eq.\eqref{Initial denisty matrix} for common qubit-noise configuration when $g=10^{-1}$ against time parameter $\tau$.}\label{different values of p}
\end{figure}

Fig.\ref{different values of p} shows the dynamics of entropic uncertainty, entropic uncertainty bound, tightness and concurrence in the bipartite entangled state when exposed to local fields and OU noise. The qualitative dynamics of the current results against different values of purity factor differ from those seen in Fig.\ref{different values of g}. In the two-qubit Werner entangled state, we find that the purity factor significantly affects the initial entanglement and level of entropic uncertainty. As seen that the relative entropic uncertainty increase proportionally as $p$ decreases, with minimum disorder in the system occurring at the upper bound of $p$ and maximum disorder at $p=0.10$. When $p>0.9$ and $p<0.1$, the $U(\tau)$ predicts minor variations between the entropic uncertainty and entropic uncertainty bound. When compared to various purity factor values, the qualitative dynamics of the current results differ from those shown in Fig.\ref{different values of g}. After a finite interval of time, the two-qubit Werner state becomes separable for all ranges of $p$. Under the OU noise, bipartite entanglement was preserved for longer intervals that shown in Refs.\cite{15, 23, 24, 27, 28, 29}.

\subsection{Entropic uncertainty relations and entanglement dynamics in independent qubit-noise configuration}
The dynamics of two qubit entropic uncertainty, entropic uncertainty bound, tightness and entanglement under the influence of classical fields characterized by OU noise are discussed in this section. The final density matrix for the IQN configuration is obtained by averaging the time evolved density matrix given in Eq.\eqref{time evolved density matrix} as:
\begin{align}
\rho_{IQN}(t)=\langle \langle \rho(\phi_1, \phi_2, t)\rangle_{\theta_1}\rangle_{\theta_2},
\end{align}
where $\chi_1 \neq \chi_2 $. The corresponding numerical form of the density matrix can be put into the following form as:
\begin{equation}
\rho_{IQN}(t)=\left(
\begin{array}{cccc}
 \frac{1+p}{4} & 0 & 0 & \frac{1}{2} e^{-4 \beta} p \\
 0 & \frac{1-p}{4} & 0 & 0 \\
 0 & 0 & \frac{1-p}{4} & 0 \\
 \frac{1}{2} e^{-4 \beta} p & 0 & 0 & \frac{1+p}{4}
\end{array}
\right)
\end{equation}
The presence of diagonal and off-diagonal components in the previous matrix indicates that the state is still entangled and coherent. As a result, time evolution limitations and noise parameter choices have a role in further restricting entanglement and promoting entropic uncertainty. Next, the analytical expressions obtained for the $U(\tau)$ and $C(\tau)$ can be given as:
\begin{align}
U(\tau)=&\frac{1}{\text{Log}[16]}e^{-4 \beta } \left(-2 p (\text{U1})+e^{4 \beta } (\text{U2}-2(\text{U3})+\text{U4})\right),\\
C(\tau)=&-\sqrt{1-p}-\frac{1}{2} \sqrt{1+p-2 e^{-4 \beta } p}+\frac{1}{2} \sqrt{e^{-4 \beta } \left(e^{4 \beta }+2 p+e^{4 \beta } p\right)}.
\end{align}
where,
\begin{align*}
U1=&2 \text{ArcTanh}\left[e^{-4 \beta } p\right]+\text{Log}\left[1+p-2 e^{-4 \beta } p\right]-\text{Log}\left[1+p+2 e^{-4 \beta } p\right],\\
U2=&-2 (1+p) \text{Log}[1+p]+(1+p) \text{Log}\left[1+p-2 e^{-4 \beta } p\right],\\
U3=&\text{Log}\left[1-e^{-4 \beta } p\right]+\text{Log}\left[1+e^{-4 \beta } p\right],\\
U4=&(1+p) \text{Log}\left[1+p+2 e^{-4 \beta } p\right].
\end{align*}

\begin{figure}[!h]
	\begin{center}
		\includegraphics[width=0.47\textwidth, height=160px]{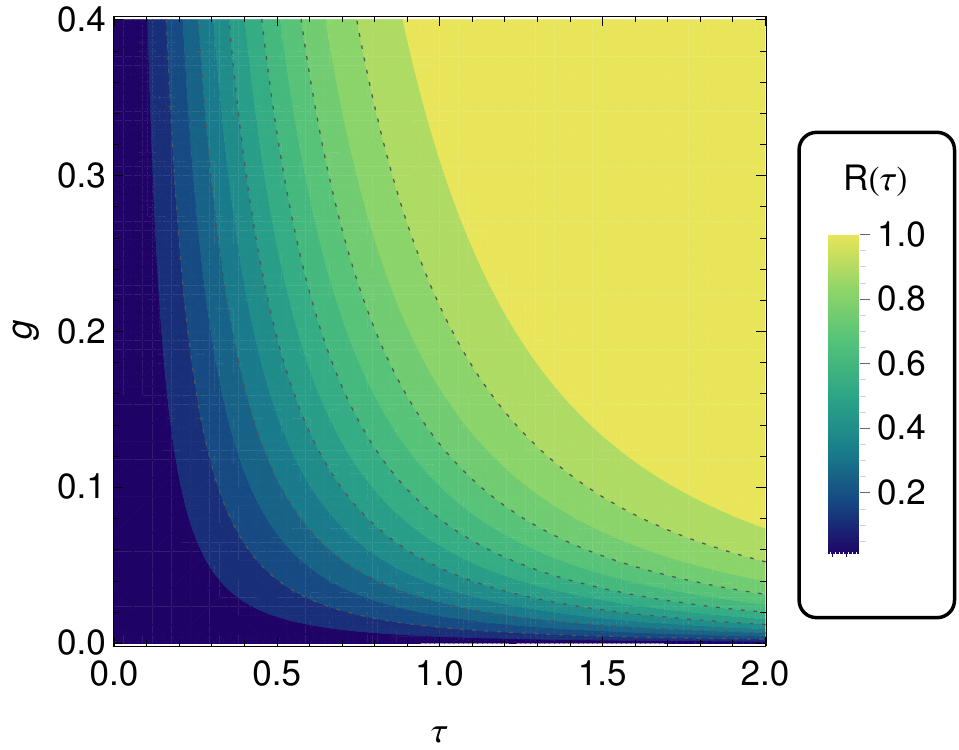}
		\put(-200,165){($ a $)} \ 
		\includegraphics[width=0.47\textwidth, height=160px]{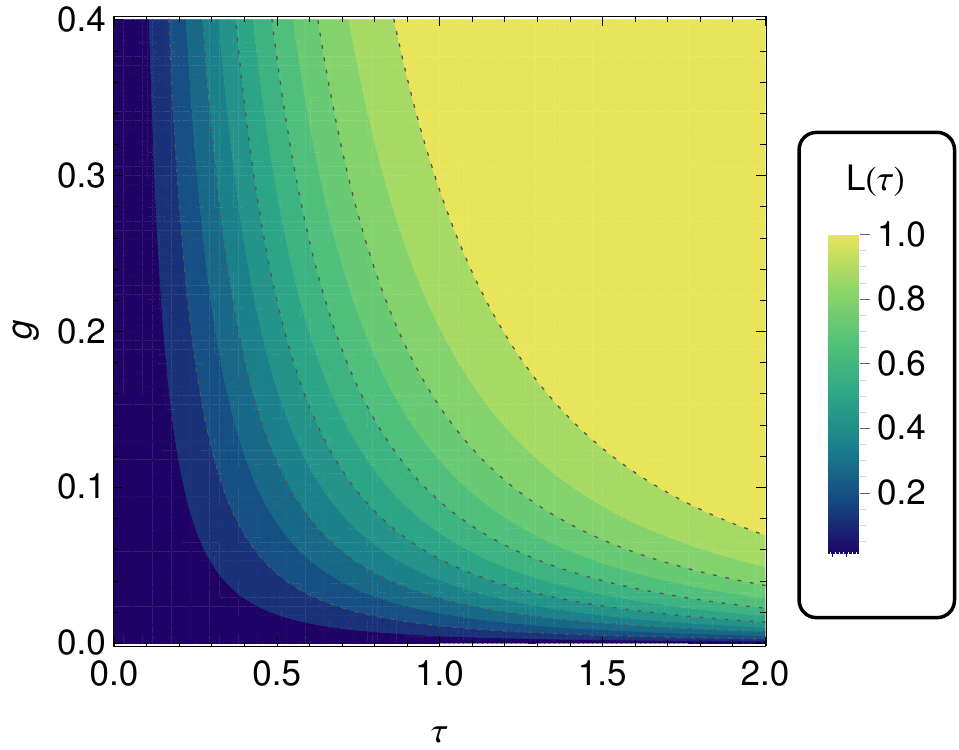}
		\put(-200,165){($ b $)} \\ 
		\includegraphics[width=0.47\textwidth, height=160px]{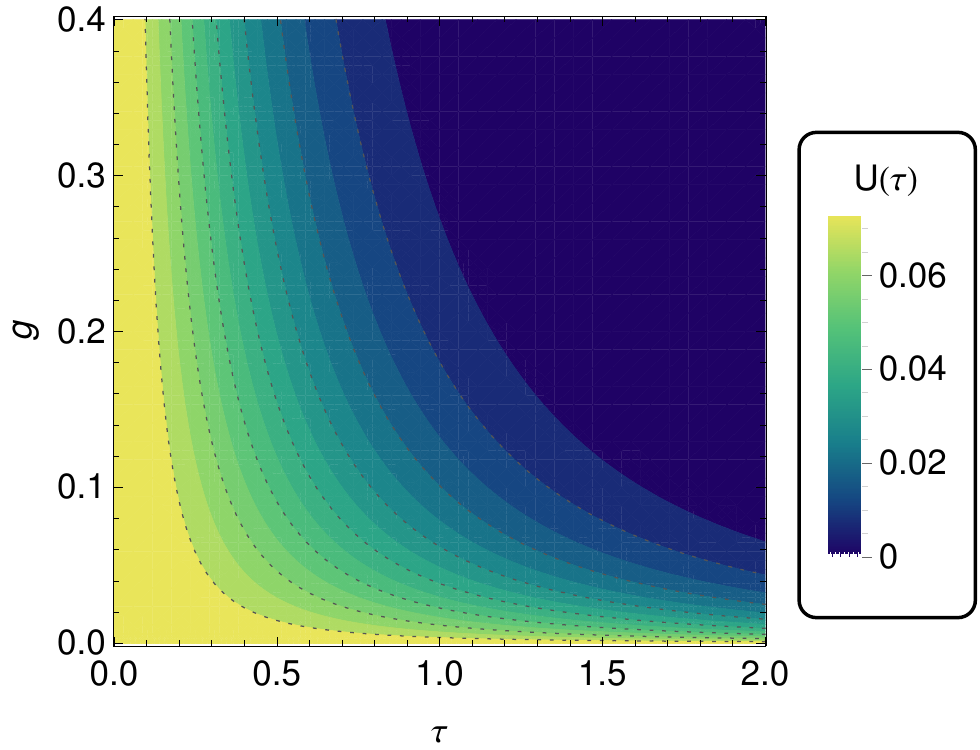}
		\put(-200,165){($ c $)}\
		\includegraphics[width=0.47\textwidth, height=160px]{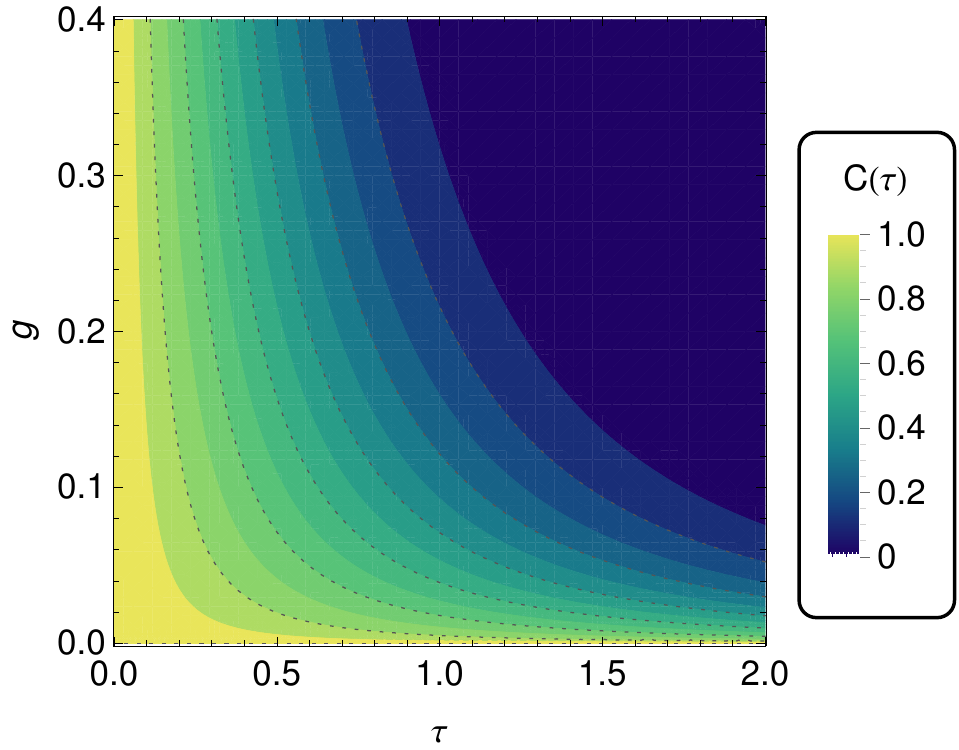}
		\put(-200,165){($ d $)} 
		\end{center}
\caption{Dynamics of $R(\tau)$ (a), $L(\tau)$ (b), $U(\tau)$ (c) and $C(\tau)$ (d) in two-qubit state, $\vert \psi \rangle=\frac{1}{\sqrt{2}}(\vert 00 \rangle+\vert 11 \rangle)$ initially prepared in the state $\rho_0$ given in Eq.\eqref{Initial denisty matrix} coupled to independent qubit-noise configuration with parameter settings: $g=0.4$, $p=1$ against time parameter $\tau$.}\label{IQN-3D}
\end{figure}
Fig.\ref{IQN-3D}  investigates the dynamics for the entropic uncertainty, entropic uncertainty bound, tightness and concurrence in two qubits coupled with two independent environments characterized by two OU noise sources. The qualitative dynamics of the entropic uncertainty and entropic uncertainty bound appear to be identical. However, the inequality remains quantitatively valid, and the two sides are not equal. It can be validated by looking at the $U(\tau)$ findings, which shows that there is a quantitative difference between the two sides, but it is lower than that seen in CQN configurations. The dynamical mappings of $L(\tau)$ and $R(\tau)$ remained growing functions of entropic uncertainty. On the other hand, $U(\tau)$ and $C(\tau)$ remained decreasing functions of tightness and entanglement. The present dynamical map of entanglement under IQN configurations appears to be more suppressed than $C(\tau)$ dynamics in CQN configurations. As a result, entanglement is better retained in CQN arrangements than in IQN configurations. The rate of entropic uncertainty rise and entanglement decrease remained proportionate and can be interpreted as that entanglement decay in the current local fields occurs latter than the entropic uncertainty rise. As a result, it may be deduced that relative entropic uncertainty causes disentanglement in the two qubits. The entropic uncertainty rises monotonically until it reaches its maximum value and finally saturates. $C(\tau)$ has a monotonic qualitative dynamical behavior that matches $L(\tau)$, $R(\tau)$, and $U(\tau)$. The new results show a lower degree of uncertainty than the tripartite entropic uncertainty explored for the three-qubit GHZ state \cite{46}. In contrast, the GHZ state remained entangled depending on the type of system-environment coupling, but the current bipartite Werner state becomes completely separable. Different types of quantum systems remained entangled depending on the type of system-environment interaction given in Refs.\cite{15, 28, 36, 37, 38}, but the current bipartite Werner state becomes completely disentangled.

\subsubsection{Explicit dynamics of the two qubits in independent qubit-noise configuration}
\begin{figure}[!h]
	\begin{center}
		\includegraphics[width=0.47\textwidth, height=160px]{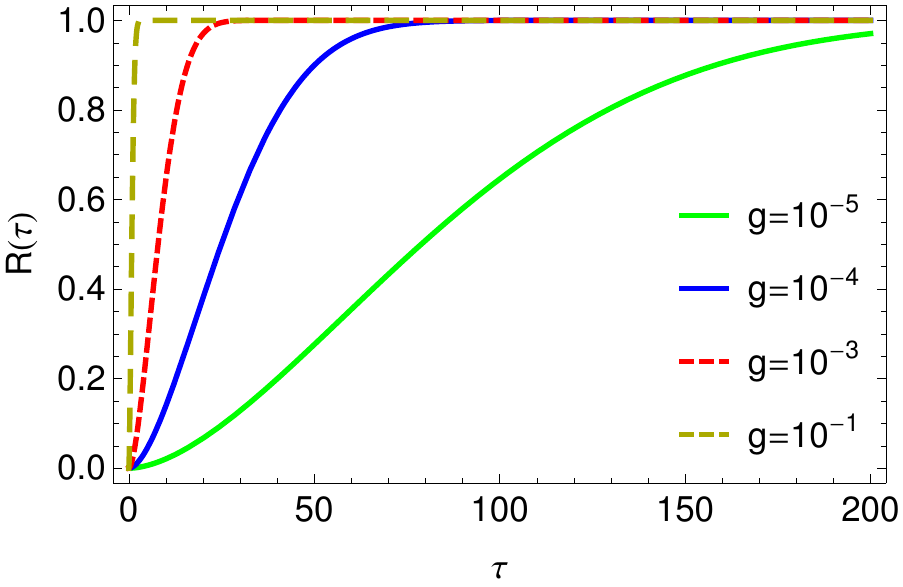}
		\put(-200,165){($ a $)} \ 
		\includegraphics[width=0.47\textwidth, height=160px]{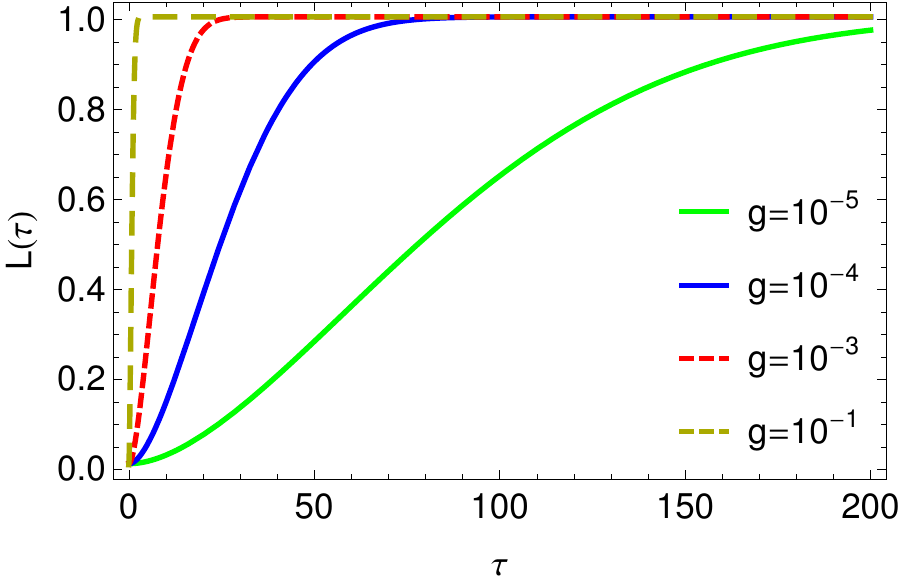}
		\put(-200,165){($ b $)} \\ 
		\includegraphics[width=0.47\textwidth, height=160px]{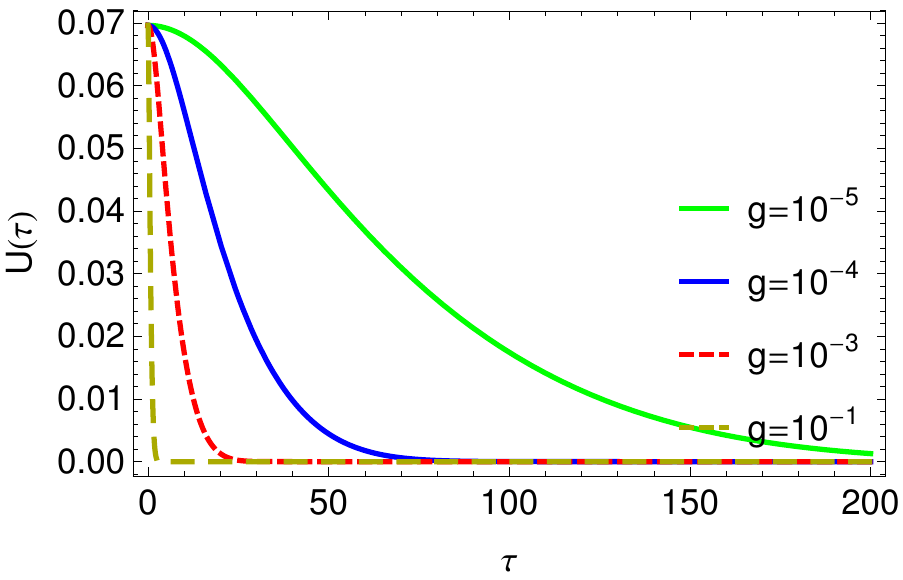}
		\put(-200,165){($ c $)}\
		\includegraphics[width=0.47\textwidth, height=160px]{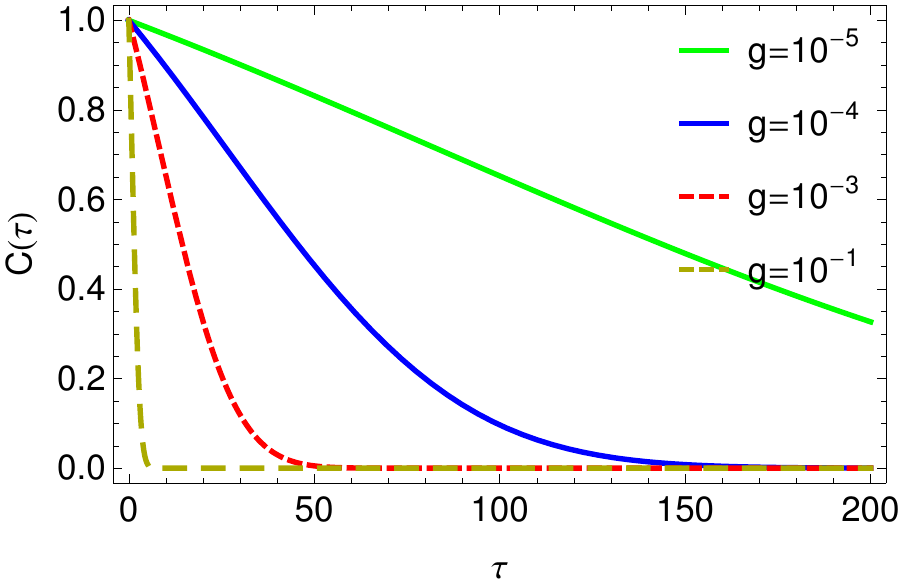}
		\put(-200,165){($ d $)} 
		\end{center}
\caption{Dynamics of $R(\tau)$ (a), $L(\tau)$ (b), $U(\tau)$ (c) and $C(\tau)$ (d) for different values of $g$ in two-qubit state, $\vert \psi \rangle=\frac{1}{\sqrt{2}}(\vert 00 \rangle+\vert 11 \rangle)$ initially prepared in the state $\rho_0$ given in Eq.\eqref{Initial denisty matrix} for independent qubit-noise configuration when $p=1$ against time parameter $\tau$.}\label{different values of g mixed}
\end{figure}

When the system is coupled with two independent classical environments, the entropic uncertainty, entropic uncertainty bound, as well as tightness and concurrence dynamics are presented in Fig.\ref{different values of g mixed}. The current findings agree with those obtained in Fig.\ref{CQN-3D}, although there are some quantitative differences. As seen from $R(\tau)$ and $L(\tau)$ functions, the present case's entropic uncertainty growing rate is slower than the CQN configuration scenario. The dynamical maps of the tripartite states under classical noises, on the other hand, demonstrate that the entropic uncertainty rise is much smaller \cite{46}. At lower $g$ values, both the $L(\tau)$ and $R(\tau)$ curves reached the ultimate saturation heights. We found the two sides deviating from each other in the curves with an insignificant narrow gap. Compared to \cite{45}, the present difference between the two sides is smaller. Even though no entropic revivals were observed in our dynamic setup, we noticed that the occurrence of revivals in the dynamical maps is due to the suppression of entropic uncertainty, as shown in \cite{49}. Although the preservation intervals are longer in this case, they are consistent with the results obtained for the two-qubit Bell's state dynamics when subjected to classical fields with static noise \cite{25}. Apart from that, as seen in Fig.\ref{different values of g}, entanglement decreases and entropic uncertainty increases faster for larger values of $g$. In contrast, we noticed that the values of the entropic uncertainty relation are lower for smaller $g$ values.

\begin{figure}[!h]
	\begin{center}
		\includegraphics[width=0.47\textwidth, height=160px]{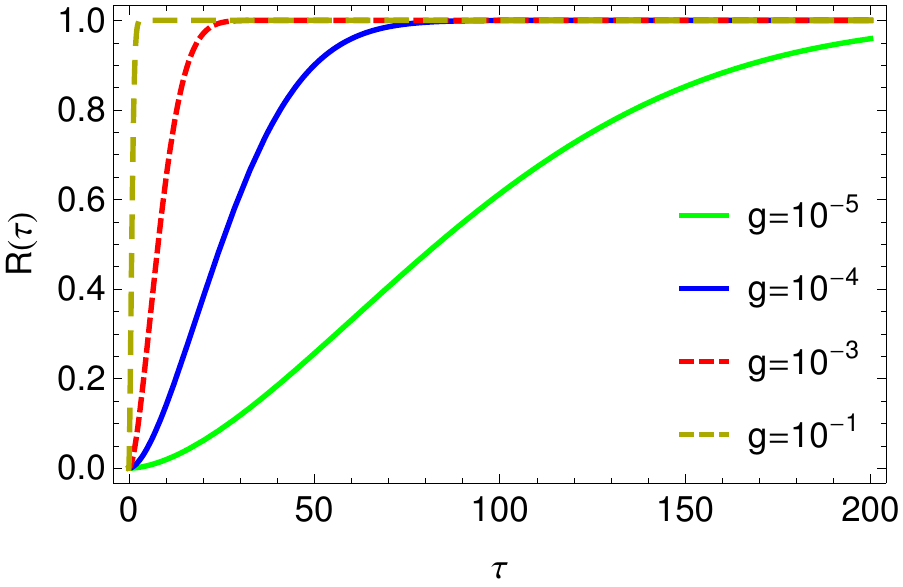}
		\put(-200,165){($ a $)} \ 
		\includegraphics[width=0.47\textwidth, height=160px]{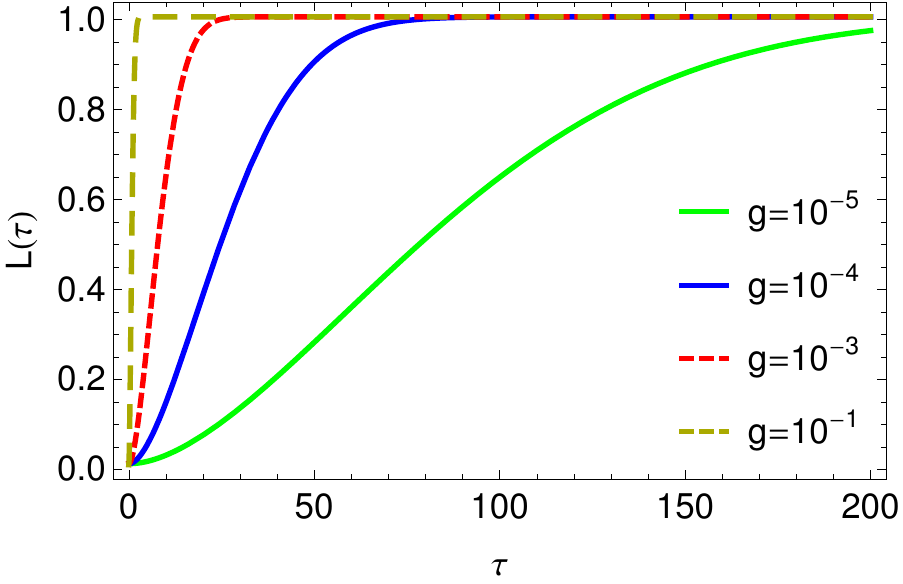}
		\put(-200,165){($ b $)} \\ 
		\includegraphics[width=0.47\textwidth, height=160px]{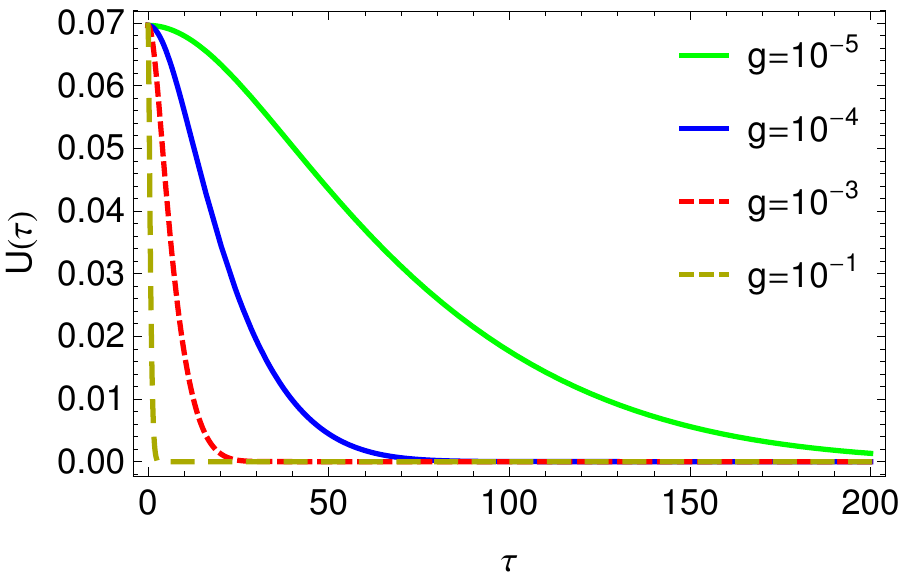}
		\put(-200,165){($ c $)}\
		\includegraphics[width=0.47\textwidth, height=160px]{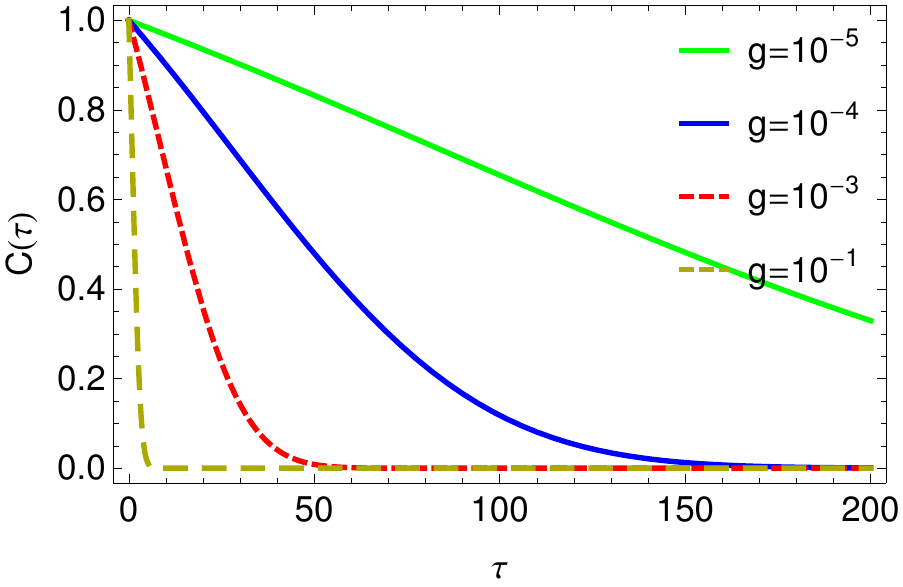}
		\put(-200,165){($ d $)} 
		\end{center}
\caption{Dynamics of $U(\tau)$ (a), $R(\tau)$ (b), $L(\tau)$ (c) and $C(\tau)$ (d) for different values of $p$ in two-qubit state, $\vert \psi \rangle=\frac{1}{\sqrt{2}}(\vert 00 \rangle+\vert 11 \rangle)$ initially prepared in the state $\rho_0$ given in Eq.\eqref{Initial denisty matrix} for independent qubit-noise configuration when $g=10^{-1}$ against time parameter $\tau$.}\label{different values of p mixed}
\end{figure}

In Fig.\ref{different values of p mixed}, we illustrate the dynamics of the two sides of entropic uncertainty relations, tightness and concurrence in the bipartite entangled state coupled to local random fields with OU noise against various purity factor values. The current findings can be traced back to different values of $p$ in Fig.\ref{different values of p}, and it can be seen that as $p$ decreases, the initial entropic uncertainty increases. Compared to the CQN set-up in the relevant values of $p$, the entropic uncertainty increase in the current case is smaller. The measurements of $R(\tau)$, and $L(\tau)$ agree, demonstrating that the initial uncertainty is lowest for $p=0.99$ and grows for all other values. Besides, the $U(\tau)$ functions has shown larger variation between the range $0.9 > p > 0.1$, and at maximum and minimum bounds of $p$, the variation between $R(\tau)$ and $L(\tau)$ functions becomes insignificant. When the difference between $R(\tau)$ and $L(\tau)$ approaches zero, the tightness curves eventually reach a minimal saturation level. The entanglement preservation duration appears to be influenced by the initial purity of the two qubits. Due to the decoherence and entropic uncertainty nature of the classical environments, the entanglement quickly fades as the initially encoded entanglement lowers. The current dynamical maps in IQN configuration offered less entropic uncertainty and a longer entanglement retention time than CQN configuration, showing that it is a good resource for practical quantum information processing. This contradicts the findings of the tripartite non-local correlations, which remained more robust and preserved in the presence of a common noise source as compared to the bipartite entangled state given in \cite{24, 28,36, 51}.

\subsection{Purity factor, degree of entropic uncertainty relations and entanglement}
The dynamics of entropic uncertainty, entropic uncertainty bound, tightness and concurrence as functions of the purity factor of the two-qubit mixed entangled state, is discussed in this section.

\begin{figure}[!h]
	\begin{center}
		\includegraphics[width=0.47\textwidth, height=160px]{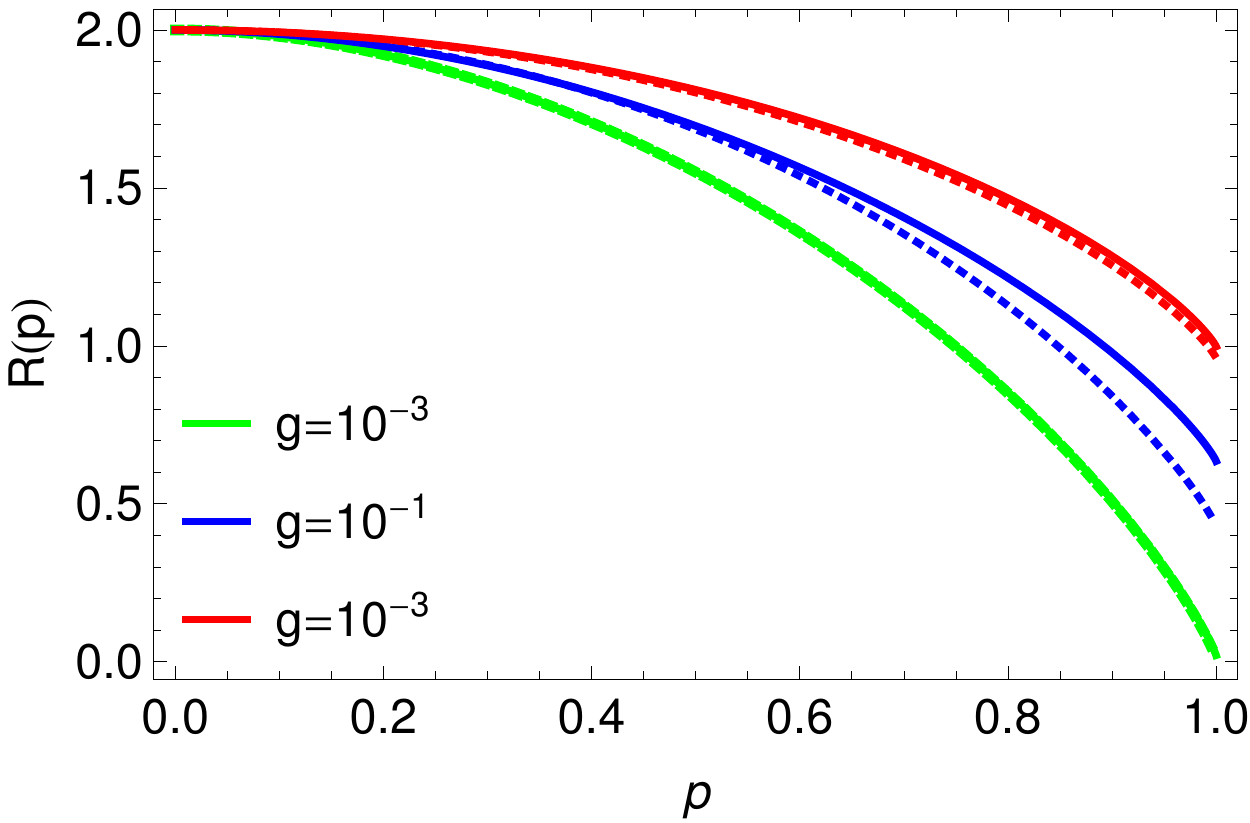}
		\put(-200,165){($ a $)} \ 
		\includegraphics[width=0.47\textwidth, height=160px]{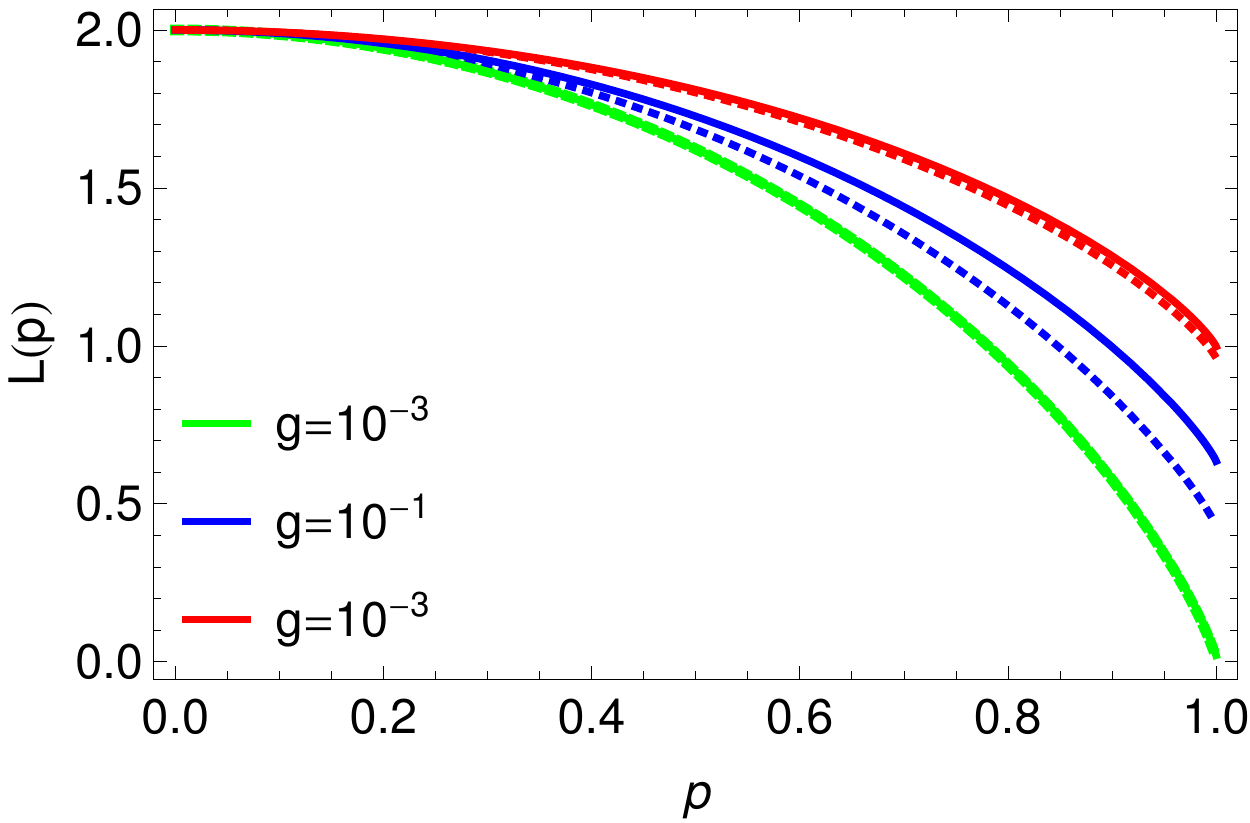}
		\put(-200,165){($ b $)} \\ 
		\includegraphics[width=0.47\textwidth, height=160px]{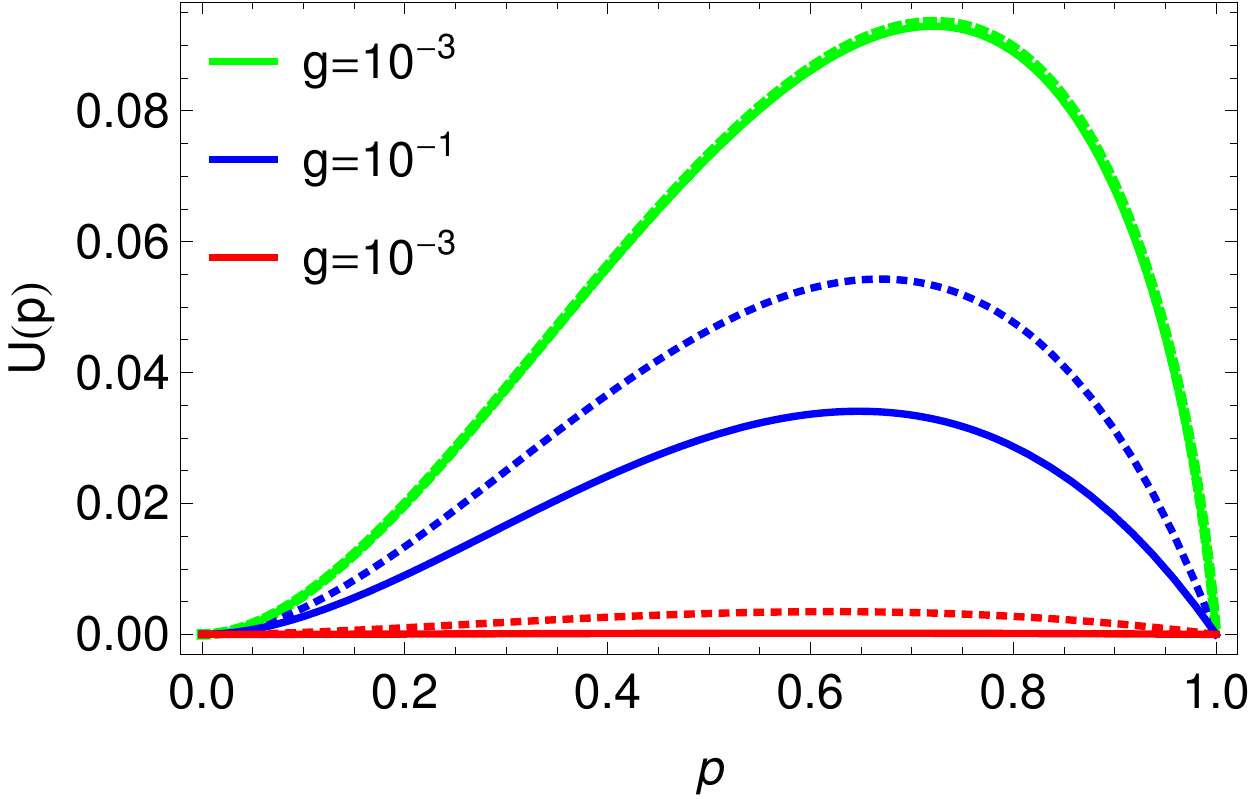}
		\put(-200,165){($ c $)}\
		\includegraphics[width=0.47\textwidth, height=160px]{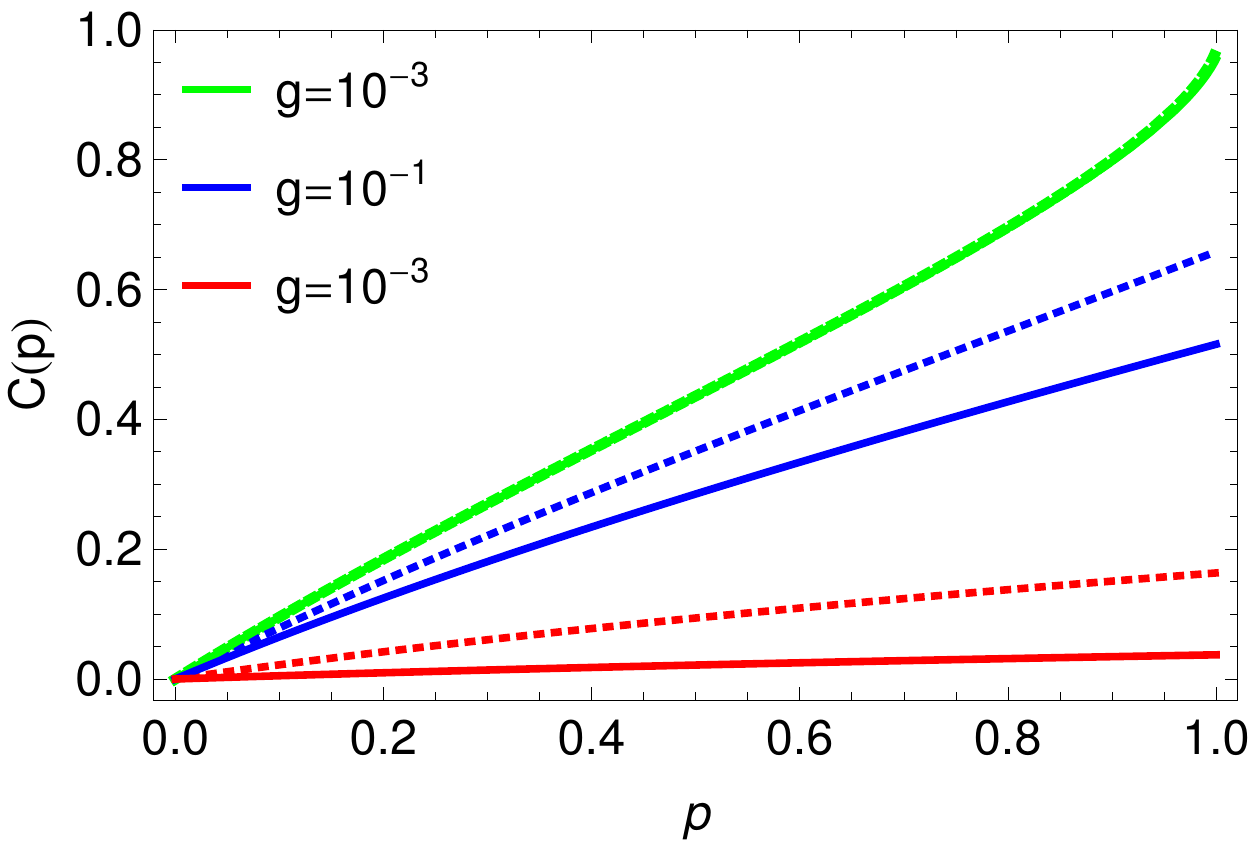}
		\put(-200,165){($ d $)} 
		\end{center}
\caption{Dynamics of $R(p)$ (a), $L(p)$ (b), $U(p)$ (c) and $C(p)$ (d) within full range $p$ in two-qubit state, $\vert \psi \rangle=\frac{1}{\sqrt{2}}(\vert 00 \rangle+\vert 11 \rangle)$ initially prepared in the state $\rho_0$ given in Eq.\eqref{Initial denisty matrix} for common qubit-noise configuration (non-dashed) and independent qubit-noise configuration (dashed) against time parameter $\tau$.}\label{p-dynamics}\end{figure}

In Fig.\ref{p-dynamics}, the entropic uncertainty, entropic uncertainty bound, tightness and concurrence are displayed as a function of purity parameter $p$. The $R(p)$ and $L(p)$ are both maximum at $p=0$ and minimum at $p=1$, according to the current results. This means that the entropic uncertainty functions reaches their maximums when the two-qubit Werner state becomes completely separable. However, in the CQN configuration, both the $L(p)$ and $R(p)$ entropies are higher than in the IQN setup. The difference between the $L(p)$ and $R(p)$ in the range $0.9 \leq p \leq 0.15$ is smaller than the top and lower limits of $p$. The results of $U(p)$ depict similar results as that of $L(p)$ and $R(p)$. At $p=0$, the state of the $C(p)$ becomes completely separable, and at $p=1$, it becomes maximally entangled. The variance in entanglement and entropic uncertainty relations becomes completely insignificant for minimum $g$ values against purity factor of the state, as seen from entropic uncertainty and concurrence functions. Compared to the CQN configuration, the entanglement in the IQN configuration appears to be better preserved. As a result, the IQN configuration may simulate quantum information processing protocols realistically.
\section{Conclusion}\label{Conclusion}
We study the dynamics of entropic uncertainty, entropic uncertainty bound, tightness and entanglement in two qubits exposed to classical fields described by the Ornstein-Uhlenbeck process. The two qubits are prepared initially in an entangled Werner state regulated by a purity factor. In addition, we consider two different system-environment coupling schemes, namely, common qubit-noise and independent qubit-noise configurations. Besides, we analyzed the entropic uncertainty, uncertainty bound, tightness and entanglement under different parameters setup to obtain optimal procedure for achieving longer entanglement and correlation time. In addition, we primarily focused on finding the relation between the entropic uncertainty and related entanglement decaying effects in the two-qubit entangled state.
\par
Our findings show that both entrropic uncertainty and entanglement in quantum systems are related, in the sense that when entropic uncertainty  rises, entanglement falls. Entropic uncertainty and entropic bound functions, both increased due to the noisy action of the classical fields in the dynamical map of two qubits. The discrepancy between the entropic uncertainty and entropic uncertainty bound is controlled primarily by the purity parameter. However, for the common qubit-noise configurations, the related uncertainty functions have shown better growth as compared to that in independent qubit-noise configuration. Besides, the difference between the two entropic uncertainty functions develops more when the noise parameter $g$ is raised. In the case of purity factor $p$, some unusual behaviour has been observed. In the ranges of $0.9 \geq p \geq 0.15$, the entropic uncertainty gap between the two sides is larger, but it becomes exceptionally small at $p>0.9$ and $p<0.15$. Our findings show that the left-hand side of the entropic uncertainty relation is more effective than the right-hand side, which is consistent with the findings given in \cite{45}. In the case of concurrence, entanglement decreases as the entropic uncertainty between the qubits increases and it remains a crucial cause for the disentanglement of the two qubits. 

The entanglement preservation intervals are controlled by intensity of the noise in classical channel(s), which decrease proportionately as $g$ rises. In contrast, the initial state entanglement and entropic uncertainty is solely dependent upon the state's purity and both are directly proportional.  

\par

Finally, under any parameter optimisation values, the two-qubit state eventually reaches separability, with no ultimate solution to avoid the corresponding disentanglement and decoherence. It's worth mentioning, however, that the phase factors of current systems can be leveraged to accomplish longer entanglement preservation via the optimal procedure, particularly when $g \leq 10^{-3}$ is employed. In contrast to the bipartite and tripartite states reported in \cite{45,15, 12, 23, 25}, entanglement in the current study has been maintained for extended periods. The entropic uncertainty relations can also be depicted similarly. Moreover, the entropic uncertainty relations and concurrence all showed a monotonic Markovian behaviour with no revivals, thus, predicting the permanent loss of information in the current classical dephasing channels.

\section{Data availability statement}
I have all data included in this manuscript.
\section{Conflicts of interests}
We authors have no conflicts of interests
\end{document}